\documentclass[journal]{IEEEtran}
\usepackage{amsmath,amsfonts}
\usepackage{algorithmic}
\usepackage{algorithm}
\usepackage{array}
\usepackage[caption=false,font=normalsize,labelfont=sf,textfont=sf]{subfig}
\usepackage{textcomp}
\usepackage{stfloats}
\usepackage{url}
\usepackage{verbatim}
\usepackage{graphicx}
\usepackage{cite}
\usepackage{booktabs}
\usepackage{multirow}
\usepackage{makecell}
\usepackage{stfloats}
\usepackage{setspace}
\usepackage{enumitem}

\begin{document}
\title{HQFNN: A Compact Quantum-Fuzzy Neural Network for Accurate Image Classification}

\author{Jianhong Yao, Yangming Guo,~\IEEEmembership{Member, IEEE}
\thanks{This paper was produced by the Jianhong Yao Group. They are in Xi'an Shaanxi, China.}
\thanks{Jianhong Yao and Yangming Guo are with School of Computer Science and Network Security College, Northwestern Polytechnical University, Xi'an Shaanxi, 710129, China.
	\\(E-mail: yjhyjhyjh@mail.nwpu.edu.cn; soblack@mail.nwpu.edu.cn)}
}

\maketitle

\begin{abstract}

Deep learning vision systems excel at pattern recognition yet falter when inputs are noisy or the model must explain its own confidence. Fuzzy inference, with its graded memberships and rule transparency, offers a remedy, while parameterized quantum circuits can embed features in richly entangled Hilbert spaces with striking parameter efficiency. Bridging these ideas, this study introduces a innovative Highly Quantized Fuzzy Neural Network (HQFNN) that realises the entire fuzzy pipeline inside a shallow quantum circuit and couples the resulting quantum signal to a lightweight CNN feature extractor. Each image feature is first mapped to a single qubit membership state through repeated angle reuploading. Then a compact rule layer refines these amplitudes, and a clustered CNOT defuzzifier collapses them into one crisp value that is fused with classical features before classification. Evaluated on standard image benchmarks, HQFNN consistently surpasses classical, fuzzy enhanced and quantum only baselines while using several orders of magnitude fewer trainable weights, and its accuracy degrades only marginally under simulated depolarizing and amplitude damping noise, evidence of intrinsic robustness. Gate count analysis further shows that circuit depth grows sublinearly with input dimension, confirming the model’s practicality for larger images. These results position the model as a compact, interpretable and noise tolerant alternative to conventional vision backbones and provide a template for future quantum native fuzzy learning frameworks.
Codes related to this work are available at httpXXXXX.

\end{abstract}
\begin{IEEEkeywords}
	Quantum Neural Network, Fuzzy Neural Network, Highly Quantized Architecture, Image Classification.
\end{IEEEkeywords}
\section{Introduction}
\IEEEPARstart{I}{mage} classification via deep convolutional neural networks (CNNs) has revolutionized fields from medical imaging to autonomous navigation by learning hierarchical feature representations directly from raw pixels. However, these deterministic models exhibit a fundamental brittleness: carefully crafted adversarial perturbations can drive error rates from under 2\% to over 80\% with imperceptible input changes \cite{goodfellow2014explaining}\cite{mkadry2017towards}, and real-world sensor noise or domain shifts similarly expose blind spots that standard training fails to address. Moreover, the common softmax confidence score is poorly calibrated as a proxy for uncertainty, leading to overconfident yet incorrect predictions, which is an unacceptable risk in safety-critical applications such as surgical assistance or self-driving vehicles. Recent advances in certification of uncertainty calibration under adversarial attack highlight the urgent need for architectures that natively quantify predictive confidence rather than layering post-hoc defenses \cite{emde2024towards}.

To this end, fuzzy neural networks (FNNs) offer a rich paradigm for uncertainty modeling and interpretability by mapping features to graded membership degrees and aggregating them via fuzzy rules. Classical FNN architectures, such as the vector-membership rule base of Angelov and Yager, demonstrate that embedding fuzzy inference within neural structures can yield robust, human-readable decision logic \cite{angelov2010simple}. More recent deep extensions, including end-to-end Neuro-Fuzzy networks for MNIST and CIFAR datasets \cite{yazdanbakhsh2019deep} and fuzzy c-means clustering layers appended to convolutional backbones \cite{yeganejou2018classification}, further validate the potential of fuzzy integration. Yet, most designs still depend on fixed-form or heuristically initialized membership functions that struggle to adapt to the intrinsic complexity of high-dimensional image manifolds, forcing an unfavorable trade-off between expressive flexibility and parameter efficiency.

In parallel, parameterized quantum circuits (PQCs) have emerged as a powerful machine-learning substrate, encoding classical inputs into exponentially large Hilbert spaces via trainable rotation and entangling gates. These models exhibit remarkable expressive power, often matching or exceeding classical counterparts with far fewer parameters \cite{benedetti2019parameterized}. Supervised classifiers leveraging quantum-enhanced feature maps have demonstrated superior separability on benchmark datasets \cite{havlivcek2019supervised}, and theoretical analyses show that PQCs can yield richer feature kernels via controlled entanglement \cite{schuld2015introduction}. However, existing quantum fuzzy hybrids invariably offload the defuzzification and rule-aggregation stages to classical layers, interrupting gradient flow across the full inference chain and diluting potential quantum speed-ups \cite{wu2024hierarchical}. 

Closing this gap demands a fully quantum-native fuzzy inference pipeline that supports end-to-end training of both membership evaluation and defuzzification, which is precisely the objective of our Highly Quantized Fuzzy Neural Network (HQFNN).

The primary contributions of this research are highlighted below:
\begin{enumerate}[label=\arabic*)]
	\item A novel neural network HQFNN is proposed. PQC structure is used to execute classical FNN’s key blocks, fuzzy membership function and defuzzification. 	Highly quantum circuit computation is implemented to give full play to the unique advantages of quantum computing in feature capture, thereby improving model performance. The whole model incorporates only a very small number of traditional neural network components. 
	
	\item Introduces a unified quantum fuzzy composition module that realizes end-to-end trainability in a single module, consisting of a fuzzifier, a single-qubit variational circuit using angle-encoded rotations to learn each feature’s fuzzy membership, and a defuzzifier, a compact entangling circuit of clustered CNOT gates that aggregates the memberships via a surrogate of the product T-norm into a single crisp output. 
	
	\item We evaluated HQFNN on several standard image‐classification benchmarks, where it consistently outperformed classical baselines in overall accuracy. To probe its resilience, we subjected the variational quantum sub-circuits to simulated noise channels and observed only marginal performance degradation, underscoring the model’s strong innate robustness. Finally, by analyzing the PQC gate count and circuit depth, we demonstrated that HQFNN’s quantum components scale efficiently, confirming the framework’s practical viability.
\end{enumerate}

The remainder of this paper is organized as follows. Section II provides an overview of prior studies in the areas of Image Classification, Fuzzy Inference, Quantum Machine Learning, and Quantum-Fuzzy Fusion for Machine Learning. We outline the HQFNN’s core architecture and algorithmic workflow in Section III. Section IV outlines the experimental setup and presents results demonstrating the model’s capabilities. Section V encapsulates our findings and proposes directions for future research.

\section{Related Work}
\subsection{Image Classification}
Early image classification began with hand-crafted features combined with classical classifiers: Scale-Invariant Feature Transform introduced local descriptors robust to scale and rotation \cite{lowe2004distinctive}, Histograms of Oriented Gradients captured gradient distributions for reliable object detection \cite{dalal2005histograms}, and Support Vector Machines provided a max-margin decision framework \cite{cortes1995support}. Spatial Pyramid Matching further improved upon bag-of-features by encoding spatial layouts with multi-level histograms \cite{lazebnik2006beyond}.

The breakthrough of deep learning emerged with LeNet-5, which demonstrated end-to-end convolutional learning on digit data \cite{lecun1998gradient}. AlexNet then showed that deeper CNNs trained on large-scale ImageNet data could vastly outperform prior models, leveraging ReLU activations, dropout, and GPU acceleration to win ILSVRC2012 \cite{krizhevsky2012imagenet}. Soon after, VGG proved that stacking small 3×3 convolutions boosted accuracy \cite{simonyan2015very}, and GoogLeNet’s Inception modules enabled multi-scale feature extraction within a single layer \cite{szegedy2015going}.

Despite these advances, training very deep networks remained challenging until ResNet introduced identity shortcuts to ease gradient flow and enable architectures hundreds of layers deep \cite{he2016residual}. EfficientNet later unified depth, width, and resolution scaling via compound coefficients to achieve state-of-the-art accuracy with remarkable efficiency \cite{tan2019efficientnet}. More recently, Vision Transformers applied self-attention to image patches and ConvNeXt modernized convolutional backbones with transformer-inspired design choices, showing that both attention-based and refined ConvNet architectures can rival or exceed prior performance \cite{dosovitskiy2021an},\cite{liu2022convnext}.

Together, these milestones illustrate the field’s evolution from handcrafted features to deep, flexible architectures. However, existing approaches still fall short of integrating quantum-native fuzzy inference into a unified, end-to-end trainable framework.

\subsection{Fuzzy Inference}
Fuzzy inference systems trace their roots to Zadeh’s introduction of fuzzy sets, which established graded membership and partial truth \cite{zadeh1965fuzzy}, and Mamdani’s first linguistic controller, demonstrating rule‐based synthesis of control actions via fuzzy if–then rules \cite{mamdani1975experiment}. The theoretical foundations were further consolidated in comprehensive treatises \cite{klir1995fuzzy}. Takagi and Sugeno then formalized parametric fuzzy models using polynomial consequents for system identification \cite{takagi1985fuzzy}, and Jang’s ANFIS merged fuzzy inference with neural networks, enabling joint learning of membership functions and rule parameters through backpropagation and least‐squares estimation \cite{jang1993anfis}.

To address higher‐order uncertainties, interval Type-2 fuzzy logic systems were proposed by Liang and Mendel, introducing efficient inference formulas and type‐reduction methods for real‐time control \cite{liang2000interval}. Subsequent simplifications by Mendel, John, and Liu \cite{mendel2006interval} reduced computational overhead, making Type-2 controllers more tractable for practical applications.

More recently, hybrid neuro-fuzzy architectures have emerged that embed fuzzy reasoning within deep learning pipelines. Yeganejou and Dick combined CNN feature extraction with deep fuzzy C-means clustering to produce interpretable classification on MNIST \cite{yeganejou2018classification}, while Yazdanbakhsh and Dick’s end-to-end deep neuro-fuzzy network integrated fuzzy inference and pooling operations as trainable layers for image recognition \cite{yazdanbakhsh2019deep}. A 2023 survey synthesizes these advances, highlighting both the scalability and explainability benefits of deep neuro-fuzzy systems \cite{talpur2022deep}.

Despite this steady progression from foundational theory through parametric, Type 2, and deep hybrid models, prior work has yet to deliver a fully quantum-native fuzzy inference pipeline that can be trained end to end.

\subsection{Quantum Machine Learning}
Early efforts to harness quantum mechanics for learning began in 1995 when Kak \cite{Kak1995} presented quantum neural computing, exploring how superposition and entanglement could augment pattern-recognition models. In 2003 Anguita et al. \cite{Anguita2003} demonstrated quantum-accelerated training of support vector machines via quadratic-programming routines, achieving logarithmic complexity improvements. Building on these foundations, in 2012 Wiebe et al. \cite{Wiebe2012} introduced a quantum algorithm for least-squares data fitting, and in 2013 Lloyd, Mohseni and Rebentrost \cite{Lloyd2013} unified supervised and unsupervised quantum learning in time logarithmic in both dataset size and dimensionality. In 2014 Rebentrost, Mohseni and Lloyd \cite{Rebentrost2014} realized a full-scale quantum support vector machine, delivering exponential speedups in both training and inference.

The variational paradigm emerged in 2018 when Mitarai et al. \cite{Mitarai2018} introduced Quantum Circuit Learning, a hybrid framework in which a low-depth parameterized circuit learns nonlinear functions by iteratively optimizing gate angles. In that same year Lloyd and Weedbrook \cite{Lloyd2018} formulated Quantum Generative Adversarial Networks, extending the adversarial training paradigm to quantum data and proving convergence to a unique fixed point with potential exponential advantage. In 2019 Cong, Choi and Lukin \cite{Cong2019} proposed Quantum Convolutional Neural Networks that embed a multiscale entanglement-renormalization ansatz into parameterized circuits for efficient recognition, and Kerenidis et al. \cite{Kerenidis2019} introduced q-means, a quantum analogue of k-means clustering with polylogarithmic scaling via QRAM-assisted inner-product estimation. Sim et al. \cite{sim2019expressibility} evaluated expressibility and entangling capability across different ansatz depths and connectivity patterns, quantifying trade-offs via KL divergence and bipartite entanglement entropy. In 2020 Pérez-Salinas et al. \cite{PerezSalinas2020} developed the Data Re-Uploading strategy, showing that repeatedly encoding classical inputs into one-qubit gates yields a universal quantum classifier without deep-circuit penalties.

As variational classifiers matured, Schuld et al. \cite{Schuld2020} designed circuit-centric quantum classifiers with polylogarithmic parameter scaling in 2020. In 2021 Pesah et al. \cite{Pesah2021} proved that QCNNs avoid barren plateaus, ensuring trainability at scale, and Du et al. \cite{Du2021} showed that depolarizing noise channels can bolster classifier robustness against adversarial perturbations. In 2022 Qu, Liu and Zheng \cite{Qu2022} extended QML to graph-structured data via temporal-spatial quantum graph convolutional networks for traffic prediction. That year, Wang et al. proposed QuantumNAS, a noise-adaptive search framework that jointly optimizes variational circuit parameters and qubit mappings under realistic noise constraints, and released TorchQuantum, a PyTorch-based, GPU-accelerated, auto-differentiable quantum circuit simulator \cite{wangQuantumNASNoiseAdaptiveSearch2022}. More recently in 2023 Skolik et al. \cite{Skolik2023} quantified the impact of hardware errors on quantum reinforcement learning and proposed measurement-efficient mitigation strategies. In addition, Xiong et al. employ parameterized quantum graph circuits within a quantum deep-reinforcement learning framework, encoding topological features in compact variational circuits to outperform classical centrality metrics with enhanced generalization \cite{xiong2025finding}.

Quantum machine learning remains constrained by fragmented inference pipelines that interrupt end-to-end training and by inefficient data-encoding schemes that cannot scale to high-dimensional inputs. Limited qubit resources further restrict circuit depth and expressiveness, while sensitivity to Noisy Intermediate-Scale Quantum (NISQ) noise goes unchecked in the absence of standardized robustness benchmarks.

\subsection{Quantum-Fuzzy Fusion for Machine Learning}
Over the past three decades, scholars have shown that the non-Boolean structure of quantum propositions admits a natural fuzzy-set interpretation. Jarosław Pykacz \cite{pykacz1992fuzzy} first illustrated how fuzzy-set theory can model quantum propositions by treating projection operators as fuzzy subsets of the state space. Building on this, Dalla Chiara and Giuntini \cite{dallachiara1996fuzzy} formalized “fuzzy quantum logics,” demonstrating that standard quantum logic is a special case of many-valued fuzzy logics, where quantum effects correspond to graded truth values. More recently, Melnichenko \cite{melnichenko2007quantum} explicitly proposed interpreting quantum logic as a fuzzy logic of fuzzy sets, unifying the mathematical frameworks and establishing a solid theoretical basis for hybrid quantum–fuzzy inference systems. Collectively, these works confirm that fusing quantum computation with fuzzy inference is not only theoretically viable but arises naturally from their shared algebraic underpinnings.

In the past two years, research on quantum–fuzzy fusion has truly begun to flourish. During this period, Tiwari et al.’s team has successively proposed two quantum-fuzzy fusion algorithms: the Quantum Fuzzy Neural Network (QFNN) \cite{tiwari2024quantum} and its federated extension, Quantum Fuzzy Federated Learning (QFFL) \cite{qu2024quantum}. QFNN employs complex-valued fuzzification via angle-encoded single-qubit circuits and a quantum defuzzifier within a seq2seq multimodal framework, achieving superior sentiment and sarcasm detection on benchmark datasets. QFFL adapts this design to distributed settings by deploying QFNNs at edge nodes and leveraging a one-shot quantum inference protocol to ensure data privacy, communication efficiency, and accelerated convergence on non-IID data. Complementary to these, Wu et al.’s QA-HFNN \cite{wu2024quantum} fuses single-qubit fuzzy membership modules with CNN-derived features for noise-robust image classification, while Jia et al.’s HGCL-QNN \cite{jia2025hierarchical} integrates fuzzy quantum encoding and hierarchical graph contrastive learning to boost multimodal sentiment analysis performance.

However, these algorithms typically offload defuzzification and rule‐aggregation to classical post‐processing stages, interrupting gradient flow and diluting potential quantum speed-ups. They also lack systematic theoretical analysis of fuzzy membership expressibility and robustness under realistic quantum noise, and remain untested on high-dimensional image-classification benchmarks. Consequently, a fully quantum-native fuzzy inference pipeline that preserves end-to-end trainability and supports rigorous theoretical and empirical evaluation remains unrealized.


\section{Methodology} 

\begin{figure*}[tp]
	\centering
	\includegraphics[width=2\columnwidth]{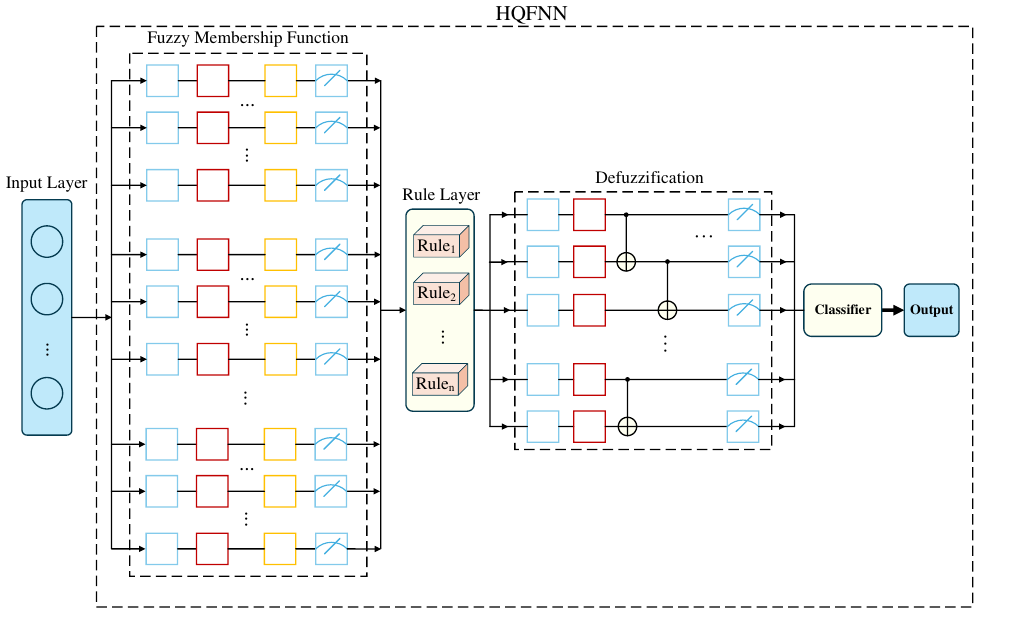}
	\caption{The architecture of HQFNN. It consists of three parts: quantum membership function, fuzzy rule layer and quantum defuzzification layer.}
	\label{fig:HQFNNFramework}
\end{figure*}

The model this paper proposes, HQFNN, is composed of three core architecture modules: quantum fuzzy membership function, rule layer and quantum defuzzification. The structure of HQFNN is illustrated in Fig. \ref{fig:HQFNNFramework}. In our framework, raw input data are first preprocessed by a lightweight CNN to produce compact feature vectors. These vectors are then fed into a dual-stream architecture where one branch extracts classical neural representations, and the other feeds into our quantum fuzzy logic module. The quantum fuzzy branch is driven by three core components: a quantum fuzzy membership function that transforms the features into complex-valued fuzzy representations, a rule layer that refines and aggregates these fuzzy outputs, and quantum defuzzification module that converts the fuzzy states into crisp quantum outputs. The outputs of these branches are fused together and finally classified by a dedicated classifier layer to yield the final predictions.


\textbf{Quantum Membership Function (QMF): }
In our proposed architecture, the Quantum Membership Function (QMF) plays a pivotal role in converting classical feature vectors into complex-valued fuzzy representations suitable for quantum processing. To formalize this component, we define the overall membership function as
\begin{equation}
	\mu_i(x_k)=f^{\mathrm{QMF}}_i(x_k)
	=\frac{\langle 0|\,U_i(x_k;\theta_i)^\dagger\,Z\,U_i(x_k;\theta_i)\,|0\rangle+1}{2},
\end{equation}
where, $x_k$ is the $k$-th input feature, $U_i(x_k;\theta_i)$ is the variational quantum circuit for the $i$-th membership function, $\theta_i$ its trainable parameter vector, and $Z$ is the Pauli-$Z$ observable. Inspired by the fuzzy logic principle that data often carry degrees of partial truth, the QMF module encodes each input feature into a quantum state whose amplitude reflects membership grades in the fuzzy domain. Specifically, the QMF leverages a multi-layer variational quantum circuit defined by trainable rotation gates (RX, RY, RZ). For completeness, we recall the unitary parametrisations of these gates,
\begin{equation}
	\begin{aligned}
		R_x(\theta) &= 
		\begin{bmatrix}
			\cos \frac{\theta}{2} & -i \sin \frac{\theta}{2} \\[5pt]
			-i \sin \frac{\theta}{2} & \cos \frac{\theta}{2}
		\end{bmatrix},\;
		R_y(\phi) = 
		\begin{bmatrix}
			\cos \frac{\phi}{2} & -\sin \frac{\phi}{2} \\[5pt]
			\sin \frac{\phi}{2} & \cos \frac{\phi}{2}
		\end{bmatrix},
		\\
		R_z(\zeta) &=
		\begin{bmatrix}
			e^{-i \zeta/2} & 0 \\[5pt]
			0 & e^{i \zeta/2}
		\end{bmatrix}.
	\end{aligned}
	\label{eq:quantum_gates}
\end{equation}
These definitions make later derivations for equation (\ref{eq:quantum_gates}) entirely self‑contained. This design allows the circuit to capture the inherent uncertainty and non-linearity in visual data while keeping the parameter count modest.
\begin{equation}
	|\psi_i\rangle = {R}_{y}(x_i + b_i)\,|0\rangle
\end{equation}
where $x_i$ is the $i$-th classical feature, $b_i$ is a learnable bias introducing global phase shifts, and $R_y(\cdot)$ denotes a rotation about the $y$-axis. Repeatedly applying ${R}_{y}(x_i + b_i)$ for all features initiates a superposition that reflects partial membership across multiple fuzzy sets.

First, classical feature vectors produced by the preprocessor are flattened (if necessary) and fed into the QMF module. Here, each feature element is angle-encoded into a qubit using parameterized rotation gates, effectively establishing the initial fuzzy superposition state: 
\begin{equation}
	|\Psi\rangle=
	\left(\bigotimes_{i=1}^{d} \text{R}_{y}(x_i + b_i)\right)\,|0\rangle^{\otimes d} 
\end{equation}
By tuning the variational parameters with backpropagation, the circuit adaptively adjusts how each feature contributes to fuzzy membership degrees. As additional layers of data re-uploading or more rotation gates are interspersed, the QMF can learn finer partitions of the feature space.

At the circuit’s end, a measurement in the Pauli-Z basis yields raw outputs ${z_i} \in [-1,\,1]$. To map them into fuzzy degrees ${\mu_i} \in [0,\,1]$, we apply:
\begin{equation}
	\mu_i = \frac{1}{2}\left(\langle Z_i\rangle + 1\right)
\end{equation}
where $\langle Z_i\rangle$ is the measured expectation of the i-th qubit’s $\sigma_z$ operator. These $\mu_i$ represent how strongly each input vector belongs to the respective fuzzy set. The subsequent layers consume these memberships to perform fuzzy inference.

Once the per-feature membership degrees $\mu$ are available, they are streamed straight into the rule layer, which treats these fuzzy values as the antecedents of learnable IF–THEN rules and aggregates them across membership functions and features. The resulting rule activations are then delivered to the quantum defuzzification module, where they are collapsed into a single crisp signal that can be concatenated with the classical feature vector before reaching the final classifier.


\textbf{Rule Layer:}
Having traversed the Quantum Membership Function, the model carries a tensor of raw membership amplitudes—${\mathbf{M}}\in\mathbb{R}^{B\times m\times d\times q}$that encodes, for every sample in the batch, how strongly each feature–qubit pair belongs to each fuzzy set. We encapsulate the entire rule‐aggregation step in one operation:
\begin{equation}
	R=\mathrm{perm}^{-1}\!\Bigl(\mathrm{ReLU}\bigl(\mathrm{Conv1d}(\mathrm{perm}(H);W_{\mathrm{rule}},b_{\mathrm{rule}})\bigr)\Bigr),
\end{equation}
where, $H\in\mathbb R^{B\times m\times d}$ collects the $m$ membership outputs over $d$ features, $\mathrm{perm}$ reorders $[B,m,d]\to[B,d,m]$, $\mathrm{Conv1d}(\cdot;W_{\mathrm{rule}},b_{\mathrm{rule}})$ is a 1-D convolution with weight $W_{\mathrm{rule}}$ and bias $b_{\mathrm{rule}}$, and $\mathrm{ReLU}$ is the element-wise activation.

To progress from these elemental degrees of truth toward a single, decision ready signal, we pass M into the Rule Layer, where the amplitudes are fused into higher order firing strengths before they are condensed by Quantum Defuzzification and ultimately interpreted by the classifier.

To fuse these values into expressive fuzzy rules we first flatten the feature–qubit pair (d,q) into a single channel index $C=dq$ and treat the membership axis as a pseudo‑temporal dimension,
\begin{equation}
	\mathbf{M}\xrightarrow{\text{reshape}}
	\tilde{\mathbf{M}}\in\mathbb{R}^{B\times C\times m},
	\qquad C=dq.
\end{equation}
A 1‑D convolution slides a learnable kernel $\mathbf{w}_{c}\in\mathbb{R}^{3}$ across that axis, generating a local firing strength
\begin{equation}
	R_{b,c,p}= \sum_{k=-1}^{1} w_{c,k}\,
	\tilde M_{b,c,p+k}+b_{c}, \qquad 1\le p\le m ,
	\label{eq:rule-layer convolution}
\end{equation}
followed by a ReLU non‑linearity $\tilde R_{b,c,p}=\max\bigl(R_{b,c,p},0\bigr)$. Because the same kernel is reused for every channel $c$, the layer implements a translation‑equivariant rule bank analogous to the graph kernels and the masked‑attention t‑norm.The equation (\ref{eq:rule-layer convolution}) can be interpreted as a differentiable surrogate of the product t‑norm.
Taking logarithms on both sides of the classical fuzzy conjunction, $\lambda_{r}= \prod_{j=1}^{d} \mu^{(r)}_{j}$, gives
\begin{equation}
	\log \lambda_{r}
	= \sum_{j=1}^{d}\log \mu^{(r)}_{j}
	\;\approx\; \mathbf{w}_{r}^{\!\top}\mathbf{m},
\end{equation}
where $\mathbf{m}$ stacks log $\mu^{(r)}_{j}$ and $\mathbf{w}_r$ acts as a first‑order Taylor kernel. Equation (\ref{eq:rule-layer convolution}) materialises this approximation in a trainable form.
After convolution we permute the tensor back and restore the original feature layout,
\begin{equation}
	\mathbf{R}= \mathrm{reshape}^{-1}\!\bigl(\tilde{\mathbf{R}}\bigr)
	\in\mathbb{R}^{B\times m\times d\times q},
\end{equation}
obtaining refined fuzzy activations that already embody IF–THEN relationships while remaining fully differentiable.
Because the convolution is confined to the comparatively short membership axis $m$, the rule layer introduces negligible overhead $\mathcal{O}(BmC)$ yet provides a powerful, gradient‑friendly vehicle for linguistic reasoning. The tensor $R$ is passed directly to the Quantum Defuzzification block, completing a coherent pipeline in which quantum encoding, fuzzy inference and classical classification are trained end‑to‑end.


\textbf{Quantum Defuzzification (QD): }
The tensor that exits the rule layer, $\mu \in \mathbb{R}^{B \times M \times d}$, is first flattened into a vector for each sample. This vector is then projected into a $q$-dimensional feature via a trainable linear layer:
\begin{equation}
	\theta \;=\;\mathrm{vec}(\mu)\,W_m \;+\; b_m \;\in\;\mathbb{R}^{B \times q}\,,
	\label{eq:Linear_information-bottleneck}
\end{equation}
where, $W_m\in\mathbb{R}^{M\!d\times q}$ is the learnable weight matrix and $b_m$ the bias.  Each component $\theta_i$ is then interpreted as a rotation angle on its qubit:
\begin{equation}
	\lvert \psi_i(0)\rangle \;=\;\mathrm{Rx}(\theta_i)\,\lvert 0\rangle\,.
\end{equation}
After encoding, qubits are entangled via triangular clusters of CNOT gates and measured in the Pauli-$Z$ basis.  The raw expectation values are normalized into $[0,1]$ by
\begin{equation}
	m_i \;=\;\frac{1}{2}\Bigl(\langle \psi_i \lvert Z \rvert \psi_i\rangle + 1\Bigr)\,.
	\label{eq:quantum membership}
\end{equation}
Finally, splitting $m=(m_1,\dots,m_q)$ into two contiguous parts $m^{(1)}\in\mathbb R^p$ and $m^{(2)}\in\mathbb R^{q-p}$, the crisp output is computed as
\begin{equation}
	y \;=\;\frac{w_{1}^{T}\,m^{(1)} \;+\; w_{2}^{T}\,m^{(2)}}{2}\,,
	\label{eq:quantum defuzz}
\end{equation}
where, $w_{1}$ and $w_{2}$ are the learned projection vectors. This produces a single interpretable scalar per sample, ready for the final classification layer.
Each component $\theta_i$ is treated as a rotation angle and prepared on its own qubit through
\begin{equation}
	\label{eq:Single‑qubit Rx encoding}
	\lvert\psi_i^{(0)}\rangle
	=R_x(\theta_i)\,|0\rangle
	=\cos\!\frac{\theta_i}{2}\,|0\rangle
	-\mathrm{i}\,\sin\!\frac{\theta_i}{2}\,|1\rangle,
\end{equation}
so that the full register state becomes $\lvert\Psi^{(0)}\rangle = \bigotimes_{i=1}^{q}\lvert\psi_i^{(0)}\rangle$. Local evidence must now be fused: a triangular pattern of CNOTs is applied inside every triple of wires, where
\begin{equation}
	\begin{aligned}
		\text{CNOT} =
		\begin{bmatrix}
			1 & 0 & 0 & 0 \\
			0 & 1 & 0 & 0 \\
			0 & 0 & 0 & 1 \\
			0 & 0 & 1 & 0
		\end{bmatrix},
	\end{aligned}
	\label{eq:CNOT_gates}
\end{equation}
followed—if at least one full cluster exists—by a wrap‑around CNOT between the last qubit of the final cluster and the first qubit of the register. After entanglement, every qubit is measured in the $Z$-basis, and the expectation value
\begin{equation}
	s_i
	=\langle\Psi^{(\mathrm{ent})}\lvert \;Z_i\;
	\lvert\Psi^{(\mathrm{ent})}\rangle\in[-1,1],
\end{equation}
is linearly mapped to a probability $m_i=(s_i+1)/2$. Gathering the results yields $\mathbf{m} =(m_1,\dots,m_q)\in[0,1]^{q}$. The classical head partitions this vector into $\mathbf{m}^{(1)}\in\mathbb{R}^{q_1}$ and $\mathbf{m}^{(2)}\in\mathbb{R}^{q_2}$ with $q_1+q_2=q$ and computes
\begin{equation}
	y^{(1)}=\mathbf{m}^{(1)}\mathbf{w}_1+\beta_1,
	\qquad
	y^{(2)}=\mathbf{m}^{(2)}\mathbf{w}_2+\beta_2,
\end{equation}
\begin{equation}
	\hat{y}= \frac{1}{2}\Bigl(y^{(1)}+y^{(2)}\Bigr),
	\label{eq:Averaged crisp output}
\end{equation}
where $\mathbf{w}_{1,2}$ and $\beta_{1,2}$ are learnable parameters. Consequently, Equations (\ref{eq:Linear_information-bottleneck})–(\ref{eq:Averaged crisp output}) therefore transform the high‑order fuzzy evidence into a single differentiable scalar while remaining fully consistent with the architecture’s linear compression layer, rotation‑based encoder, clustered CNOT entanglement pattern, and dual‑head regression block.
The crisp value $\hat{y}$ (vector‑valued when multiple classes are required) is finally fed to a lightweight fully connected classifier, which either augments or replaces residual classical features. This brief final stage translates the quantum‑resolved signal into logit space, enabling end‑to‑end optimisation of the entire hybrid pipeline without disrupting gradient flow.

\begin{figure*}[b]
	\centering
	\includegraphics[width=2\columnwidth]{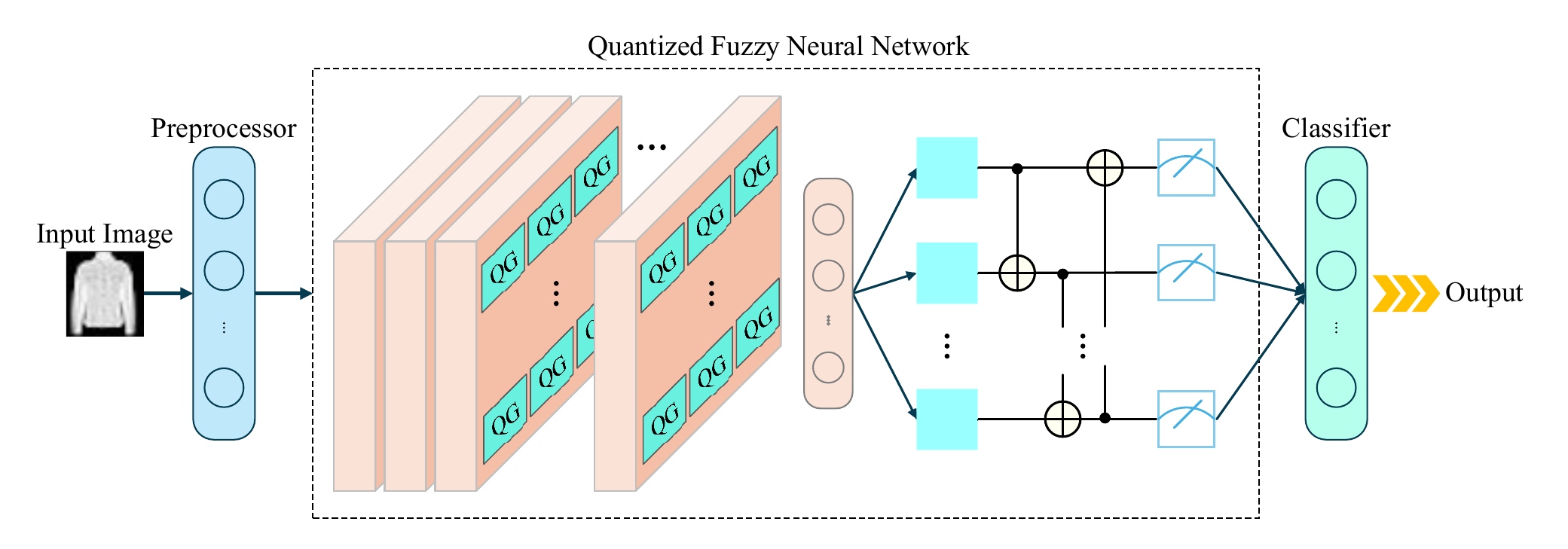}
	\caption{Visualization workflow of the HQFNN model. QG represents a trainable quantum gate.}
	\label{fig:Data_Flow}
\end{figure*}

To make the dataflow through HQFNN concrete, Algorithm~\ref{alg:hqfnn-forward} walks through a single forward pass for an entire mini‑batch. Starting with raw images, it traces how features are pre‑processed, converted into complex‑valued fuzzy memberships, aggregated by the rule layer and defuzzified into a crisp quantum signal before classification. The step‑by‑step outline also clarifies the tensor shapes and notational shortcuts used in the mathematical derivations above.
The detailed image-processing workflow of the proposed model is depicted in Fig. \ref{fig:Data_Flow}.
\begin{algorithm}[t]  
	\caption{Forward Pass of the HQFNN}
	\label{alg:hqfnn-forward}
	\begin{algorithmic}[1]  
		\REQUIRE A mini–batch $\mathcal{X}=\{\mathbf{x}_b\}_{b=1}^{B}$,
		trained parameters $\Theta=\{\theta_{\text{CNN}},\theta_{\text{QMF}},
		\theta_{\text{rule}},\theta_{\text{QD}},\theta_{\text{clf}}\}$.
		\ENSURE Logits $\mathbf{y}\in\mathbb{R}^{B\times C}$.
		\STATE \textbf{Preprocessing:}
		$\mathbf{v}_b \!\leftarrow\! \text{CNN}_{\theta_{\text{CNN}}}(\mathbf{x}_b)$
		\hfill\COMMENT{$\mathbf{v}_b\in\mathbb{R}^{d}$}
		\STATE \textbf{Quantum Membership Function (QMF):}
			\STATE \hspace{1em} Encode $\mathbf{v}_b$ via $d$ angle‑encodings $R_y(\cdot)$ on qubits $q_i$  
			\STATE \hspace{1em} Apply $L_q$ variational layers $\mathcal{U}_{\theta_{\text{QMF}}}$
			\STATE \hspace{1em} Measure $\sigma_z$\;$\Rightarrow$ fuzzy tensor $\boldsymbol{\mu}_b\!\in\![0,1]^{M\times d}$
			\STATE \textbf{Rule Layer:}
		$\mathbf{R}_b \!\leftarrow\!
		\text{ReLU}\;\!\big(\text{Conv1D}_{\theta_{\text{rule}}}
		\!\big(\text{reshape}(\boldsymbol{\mu}_b)\big)\big)$
		\STATE \hspace{1em} Restore shape $\mathbf{R}_b\!\in\!\mathbb{R}^{M\times d}$
		\STATE \textbf{Quantum Defuzzification:}
			\STATE \hspace{1em} Flatten $\mathbf{R}_b$ and project: $\boldsymbol{\theta}_b = \mathbf{R}_b \mathbf{W}_m + \mathbf{b}_m$
			\STATE \hspace{1em} Prepare $\{R_x(\theta_{b,i})\}_{i=1}^{q}$, entangle in $q/3$ CNOT clusters
			\STATE \hspace{1em} Measure $\sigma_z$ and map to $m_{b,i}=(s_{b,i}+1)/2$ $\Rightarrow$ $\mathbf{m}_b$
			\STATE \hspace{1em} Split $\mathbf{m}_b$ into $\mathbf{m}^{(1)}_b,\mathbf{m}^{(2)}_b$; compute $y_b$ via~Eq.\,(\ref{eq:quantum membership})-(\ref{eq:quantum defuzz})
			\STATE \textbf{Classifier:} 
		$\mathbf{y}_b \!\leftarrow\! \text{FC}_{\theta_{\text{clf}}}(y_b)$
		\RETURN $\mathbf{y}=\{\mathbf{y}_b\}_{b=1}^{B}$
	\end{algorithmic}
\end{algorithm}

\begin{figure*}[t]
	\centering
	\includegraphics[width=2\columnwidth]{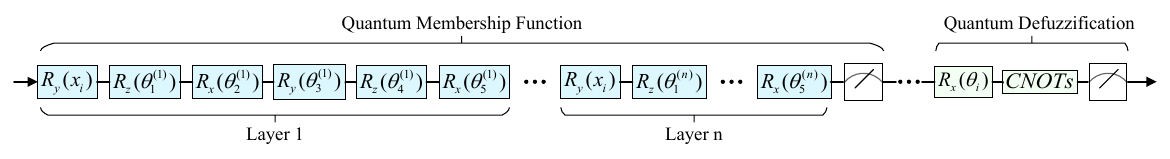}
	\caption{Quantum circuit diagram of the whole quantum block.}
	\label{fig:Quantum circuit}
\end{figure*}

While Algorithm~\ref{alg:hqfnn-forward} captures inference, Algorithm~\ref{alg:train-hqfnn} summarises the full optimisation loop that learns HQFNN’s parameters. It shows how parameter‑shift gradients from the quantum layers are back‑propagated alongside classical gradients, how Adam and a piece‑wise learning‑rate schedule are applied, and when validation checks are performed. Presenting this routine in pseudocode keeps the main text focused on the architectural ideas while still providing all implementation details required for reproducibility.

\begin{algorithm}[t] 
	\caption{End‑to‑end training routine of HQFNN}
	\label{alg:train-hqfnn}
	\begin{algorithmic}[1]

		\REQUIRE Training set $\mathcal{D}_{\text{train}}$, validation set $\mathcal{D}_{\text{val}}$, epochs $T$, batch size $B$, learning rate $\eta_0$, milestones $\mathcal{M}$.

		\ENSURE Optimised parameters $\Theta^\star$.
		
		\STATE Initialise parameters $\Theta$
		\STATE $\eta \leftarrow \eta_0$ \COMMENT{Adam optimiser, parameter–shift gradients}
		
		\FOR{$t = 1$ \TO $T$}
		\FORALL{mini‑batches $(x,y) \in \mathcal{D}_{\text{train}}$ of size $B$}
		\STATE $h \leftarrow \text{Preprocessor}(x)$
		\STATE $\mu \leftarrow \text{QMF}(h)$
		\STATE $\tilde{\mu} \leftarrow \text{RuleLayer}(\mu)$
		\STATE $z \leftarrow \text{QD}(\tilde{\mu})$
		\STATE $\hat{y} \leftarrow \text{Classifier}(z)$
		\STATE $\mathcal{L} \leftarrow \text{CrossEntropy}(\hat{y},y)$
		\STATE Compute $\nabla_{\Theta}\mathcal{L}$ by back‑propagation  
		\STATE $\Theta \leftarrow \text{AdamUpdate}(\Theta,\nabla_{\Theta}\mathcal{L},\eta)$
		\ENDFOR
		\IF{$t \in \mathcal{M}$}
		\STATE $\eta \leftarrow 0.1 \times \eta$
		\ENDIF
		\STATE Evaluate on $\mathcal{D}_{\text{val}}$ and save best $\Theta^\star$
		\ENDFOR
		\STATE \textbf{return} $\Theta^\star$
	\end{algorithmic}
\end{algorithm}

\section{Experiment}
\subsection{Experimental Settings}

\textit{I. Experimental Configuration:} Utilizing Debian 12 (Linux) on an x86-64 architecture, the classical networks and quantum circuits were simulated via GPU acceleration. Experimental hardware specifications are provided in detail in Table \ref{tab:Device information}. The quantum circuits and environments were developed with Torch Quantum\cite{wangQuantumNASNoiseAdaptiveSearch2022}, while the classical neural networks were constructed using PyTorch. Model training employed cross-entropy as the performance metric.

\textit{II. Evaluation Metrics:} In image classification tasks, model performance is evaluated using accuracy, precision, recall and F1-score. Accuracy reflects the proportion of correctly classified samples, precision measures the correctness of positive predictions, recall indicates the ability to identify all relevant instances and F1-score balances precision and recall when class distributions are uneven or error types vary. This study employs these four metrics for a comprehensive assessment of HQFNN performance..

\textit{III. Dataset Details:} To assess the model’s performance, we conduct experiments on the MNIST, JAFFE (Japanese Female Facial Expression), and Fashion-MNIST datasets.

The MNIST dataset, released by the National Institute of Standards and Technology\cite{digit-recognizer}, contains grayscale images (28×28 pixels) of handwritten digits and is commonly utilized in evaluating image processing methods.

Fashion-MNIST comprises 70,000 grayscale images (28×28 pixels) spanning 10 clothing categories\cite{xiao2017fashionmnistnovelimagedataset}, of which 60,000 are used for training and 10,000 for testing.

Dirty-MNIST merges the standard MNIST digits with the Ambiguous-MNIST set, yielding 60 000 labeled examples for training \cite{mukhoti2021deterministic}. The Ambiguous-MNIST component contains 6 000 synthetically generated images, each annotated with ten different labels to capture varying levels of uncertainty.

The JAFFE dataset contains 213 images capturing seven distinct facial expressions from ten Japanese female subjects \cite{lyons2021excavatingaireexcavateddebunking}, \cite{670949}. Among these seven expressions, one is neutral, and the other six represent various facial emotions. Each image is sized 256×256 pixels. Before use, the images undergo preprocessing to extract and isolate the facial expression regions.

Chest radiographs paired with lung segmentation masks make up the COVID-19 dataset \cite{chowdhury2020can}, featuring four classes—COVID-19 (COV), lung opacity, viral pneumonia, and normal—and serving to differentiate lung disease types through combined analysis of X-ray images and their corresponding masks.

\textit{IV. Hyperparameters:} The experimental hyperparameter configuration encompasses both the classical neural network parameters and those governing the quantum circuits. Table \ref{HYPERPARAMETER SETTING} presents the hyperparameter configuration used for HQFNN.
\begin{table}[htbp] 
	\caption{DEVICE INFORMATION. \label{tab:Device information}}
	\centering
	\setlength{\tabcolsep}{14pt}
	\begin{tabular}{lll}
		\toprule
		GPU & CPU & RAM \\
		\midrule
		{\makecell[l]{NVIDIA® GeForce \\ RTX™ 5080}} & {\makecell[l]{Intel® Core™ \\ i9-14900KF}} & 64GB \\
		\bottomrule
	\end{tabular}
\end{table} 

\begin{table}[htbp]
	\caption{HYPERPARAMETER SETTING DETAILS. \label{HYPERPARAMETER SETTING}}
	\centering
	\setlength{\tabcolsep}{25pt}
	\begin{tabular}{l c}
		\toprule
		Hyperparameter       & Value   \\
		\midrule
		 Batch size          & 500     \\
		 Learning rate       & 0.001   \\
		 Activation function & ReLU    \\
		 QMF circuit layer   & 4       \\
		 QD qubits           & 6       \\
		 QMF encoding gate   & $R_y$   \\
		 QD encoding gate    & $R_x$   \\
		 Measurement         & Pauli-Z \\
		\bottomrule
	\end{tabular}
\end{table}

\subsection{Comparative Analysis of Baseline and Proposed Algorithms}

To benchmark HQFNN, we evaluate its performance on MNIST and Fashion-MNIST against six representative baselines, including Quantum Convolutional Neural Network (QCNN) which replaces classical convolutional filters with parameterized quantum kernels to capture nonlocal feature correlations, Pattern-Parallel Hierarchical Fuzzy Neural Network (PP-HFNN) which implements a multi-stage fuzzy inference pipeline with parallel pattern processing and hierarchical aggregation, Quantum Fuzzy Federated Learning (QFFL) which embeds quantum-enhanced fuzzy inference within a federated learning framework, Quantum Neural Network (QNN) which uses layered rotation and entangling gates to form a parameterized quantum classifier, Quantum Fuzzy Neural Network (QFNN) which integrates quantum-based membership functions and defuzzification into a classical fuzzy pipeline, and Quantum-Assisted Hybrid Fuzzy Neural Network (QA-HFNN) which couples deep classical feature extractors with quantum fuzzy reasoning. These baselines encompass both purely quantum approaches and quantum-fuzzy fusion paradigms, offering a comprehensive benchmark for classification accuracy.

\begin{figure}[t]
	\centering
	\includegraphics[width=1\columnwidth]{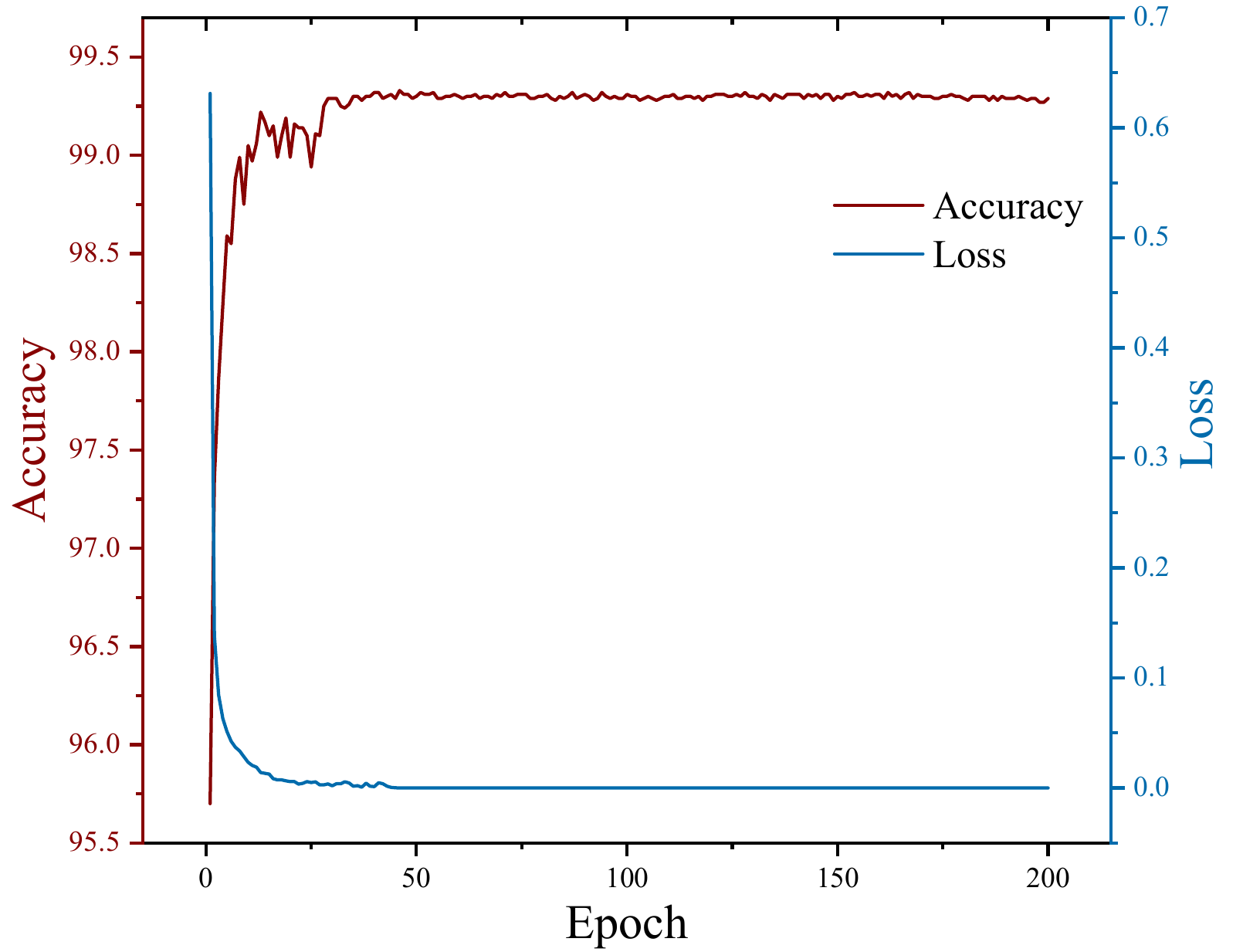}
	\caption{Training convergence of HQFNN on MNIST.}
	\label{fig:AccLossEpoch}
\end{figure}

Fig.~\ref{fig:AccLossEpoch} plots HQFNN’s training convergence on MNIST over 200 epochs. The model learns extraordinarily fast in the early stages: the training loss plunges from 0.631 at epoch~1 to 0.042 at epoch~6, while accuracy jumps from 95.70\% to 98.55\%. By epoch~5, accuracy has already cleared 98.50\%, and by epoch~10 the loss has dipped below 0.03 as accuracy surpasses 99.00\%. Thereafter, the loss drifts into the $10^{-3}$ regime—reaching roughly 0.00195 by epoch~30 and about $2.1\times10^{-5}$ by epoch~50, while accuracy simultaneously settles in the 99.31\%--99.40\% range.
Beyond this point the curves flatten, indicating the model has reached its capacity on this task. Further accuracy gains would require changes in architecture, regularization, or data augmentation. This plateau reflects the limits of the current HQFNN configuration and suggests that additional progress depends on structural refinements or new data rather than more training.

Following convergence analysis, we assessed classification on MNIST and Fashion-MNIST using a unified preprocessing pipeline—tensor conversion, [0,1] scaling and normalization by dataset statistics. HQFNN integrates a two-layer CNN feature extractor with a three-block single-qubit QNN membership module, merging fuzzy degrees and deep features through an end-to-end fusion layer. As shown in TABLE~\ref{tab:Performance}, this design delivers around 99.4\% accuracy on MNIST and 92.8\% on Fashion-MNIST, outperforming classical DNN, fuzzy-enhanced and quantum-only baselines. Its gains in precision, recall and F1-score further confirm the strength of our quantum-fuzzy fusion strategy.

TABLE~\ref{tab:Performance on Various Datasets} simply presents the accuracy of HQFNN on other datasets, they are Dirty-MNIST, JAFFE and COVID-19. And the models used as benchmarks are the most present and representative.
On Dirty-MNIST, the result of HQFNN is statistically indistinguishable from that of QA-HFNN, indicating no significant improvement.
For the COVID-19 containing real lung X-ray images, our model achieved an accuracy of 91.76\%, which is 2.04\% and 0.83\% higher than QFNN and QA-HFNN respectively. 
Similarly, in the dataset with real face images, JAFFE, we also achieved 3.77\% and 1.04\% improvement compared to the benchmarks. These results demonstrate that our quantum-fuzzy fusion mechanism generalizes effectively across diverse classification challenges.

The improved classification accuracy of HQFNN stems from its fully differentiable quantum–fuzzy inference pipeline and seamless integration with classical feature extraction. First, each input feature is encoded into a single-qubit state by parameterized rotations \(R_{x}\), \(R_{y}\), and \(R_{z}\). A cascade of multiple identical such rotation layers, each introducing three trainable angles, produces a high-resolution membership representation that flexibly adapts to complex uncertainty in image data. Next, these learned membership degrees are aggregated by a fuzzy-rule convolution module, which captures interactions across all fuzzy sets and feature channels and reduces the output back to the original feature dimensionality while preserving higher-order correlations. In the defuzzification stage, the aggregated tensor is linearly projected to a single scalar, which is then encoded across multiple qubits, entangled via clustered CNOT operations, and measured. The measurement outcomes are subsequently re-scaled and averaged to yield a crisp quantum-informed feature. Finally, concatenating this quantum-derived scalar with the classical feature embedding and passing the result through a compact two-layer classifier enables HQFNN to leverage both quantum-enhanced membership modeling and deep learned representations, resulting in superior discriminative performance these datasets.

\begin{table*}[htbp]
	\renewcommand{\arraystretch}{1.3} 
	\centering
	\caption{Performance Comparison with Baselines on MNIST and Fashion-MNIST.}
	\begin{tabular}{lccccccccc}
		\hline
		\multirow{2}{*}{Model} & \multicolumn{4}{l}{MNIST}                             && \multicolumn{4}{l}{Fashion-MNIST} \\
		\cline{2-5}\cline{7-10}
		               & Acc           & Precision     & Recall        & F1-Score      && Acc           & Precision     & Recall        & F1-Score       \\
		\hline
        QCNN           &  0.9023       &   0.8756      &   0.8533      &   0.8760      &&  0.8723       &   0.8701      &   0.8715      &   0.8710       \\
		PP-HFNN        &  0.8933       &   0.8255      &   0.8913      &   0.8516      &&  0.8521       &   0.8500      &   0.8512      &   0.8506       \\
		QFFL           &  0.9885       &   0.9884      &   0.9886      &   0.9885      &&  0.9052       &   0.9048      &   0.9051      &   0.9050       \\		
		QNN            &  0.9870       &   0.9855      &   0.9850      &   0.9852      &&  0.8830       &   0.8805      &   0.8822      &   0.8813       \\
		QFNN           &  0.9587       &   0.9585      &   0.9583      &   0.9578      &&  0.8992       &   0.8985      &   0.8990      &   0.8987       \\
		QA-HFNN        &  0.9913       &   0.9913      &   0.9912      &   0.9912      &&  0.9070       &   0.9065      &   0.9069      &   0.9061       \\
		\textbf{HQFNN} &\textbf{0.9940}&\textbf{0.9939}&\textbf{0.9938}&\textbf{0.9939}&&\textbf{0.9281}&\textbf{0.9279}&\textbf{ 0.9281}&\textbf{0.9280} \\
		\hline
	\end{tabular}
	\label{tab:Performance}
\end{table*}

\begin{table*}[htbp]
	\caption{Performance Comparison of the Proposed Model on Various Datasets.}
	\label{tab:Performance on Various Datasets}
	\centering
	\setlength{\tabcolsep}{18pt}
	\begin{tabular}{l  c c c}
		\toprule
		MODEL          & COVID-19         & JAFFE            & Dirty-MNIST           \\
		\midrule							                                      		                                    
		QFNN           & 0.8972           & 0.9057           &  0.7862              \\
		QA-HFNN        & 0.9093           & 0.9330           &  0.8400              \\
		\textbf{HQFNN} &\textbf{0.9176}   &\textbf{0.9434}   & \textbf{0.8400}       \\
		\bottomrule
	\end{tabular}
\end{table*}
\subsection{Robustness Evaluation}

\begin{table}[htbp]
	\caption{The fidelity of noisy quantum circuit.}
	\label{tab:The fidelity on noise}
	\centering
	\setlength{\tabcolsep}{12pt}
	\begin{tabular}{c c c c c}
		\toprule
		P         & AD      & DP      & BF      & PF      \\
		\midrule                                            
		0.01      & 0.9961  & 0.9746  & 0.9849  & 0.9946  \\
		0.05      & 0.9803  & 0.8798  & 0.9255  & 0.9729  \\
		0.10      & 0.9601  & 0.7757  & 0.8540  & 0.9461  \\
		\bottomrule
	\end{tabular}
\end{table}

\begin{figure*}[!t]
\centering
\subfloat[Amplitude Damping]{\includegraphics[width=2.7in]{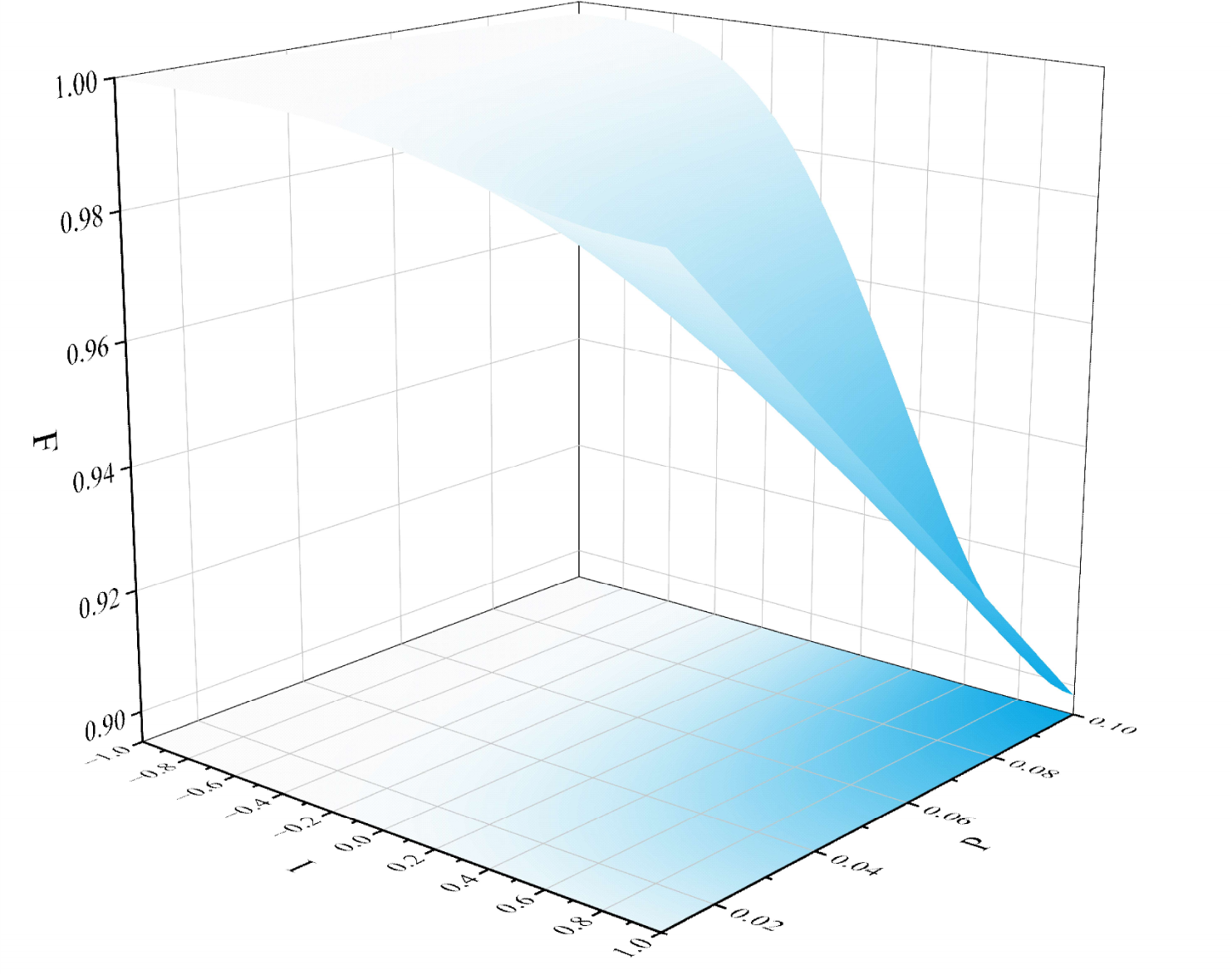}
\label{fig:AD}}
\hfil
\subfloat[Depolarizing]{\includegraphics[width=2.7in]{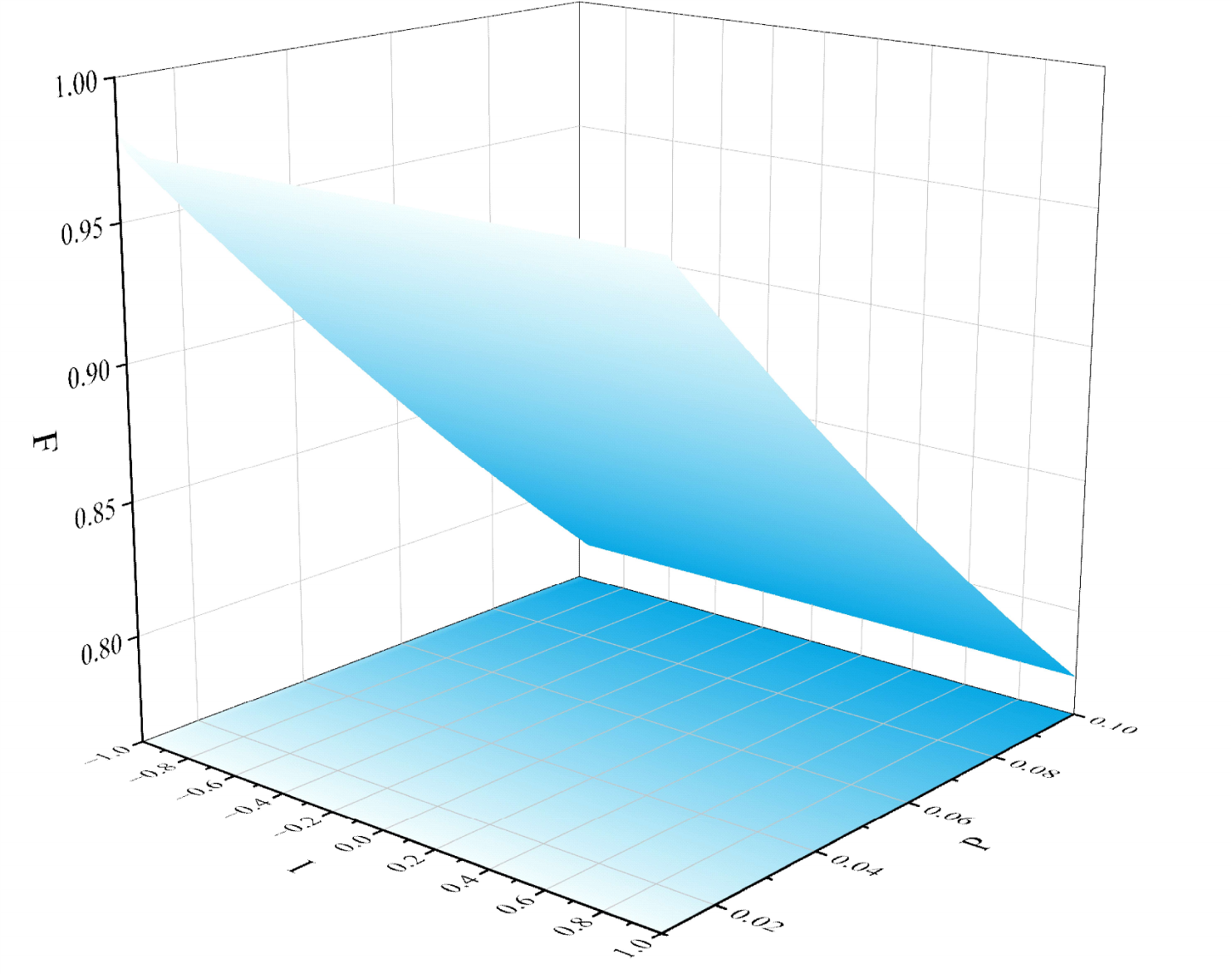}
\label{fig:DP}}
\vspace{3ex}
\centering
\subfloat[Bit Flip]{\includegraphics[width=2.93in]{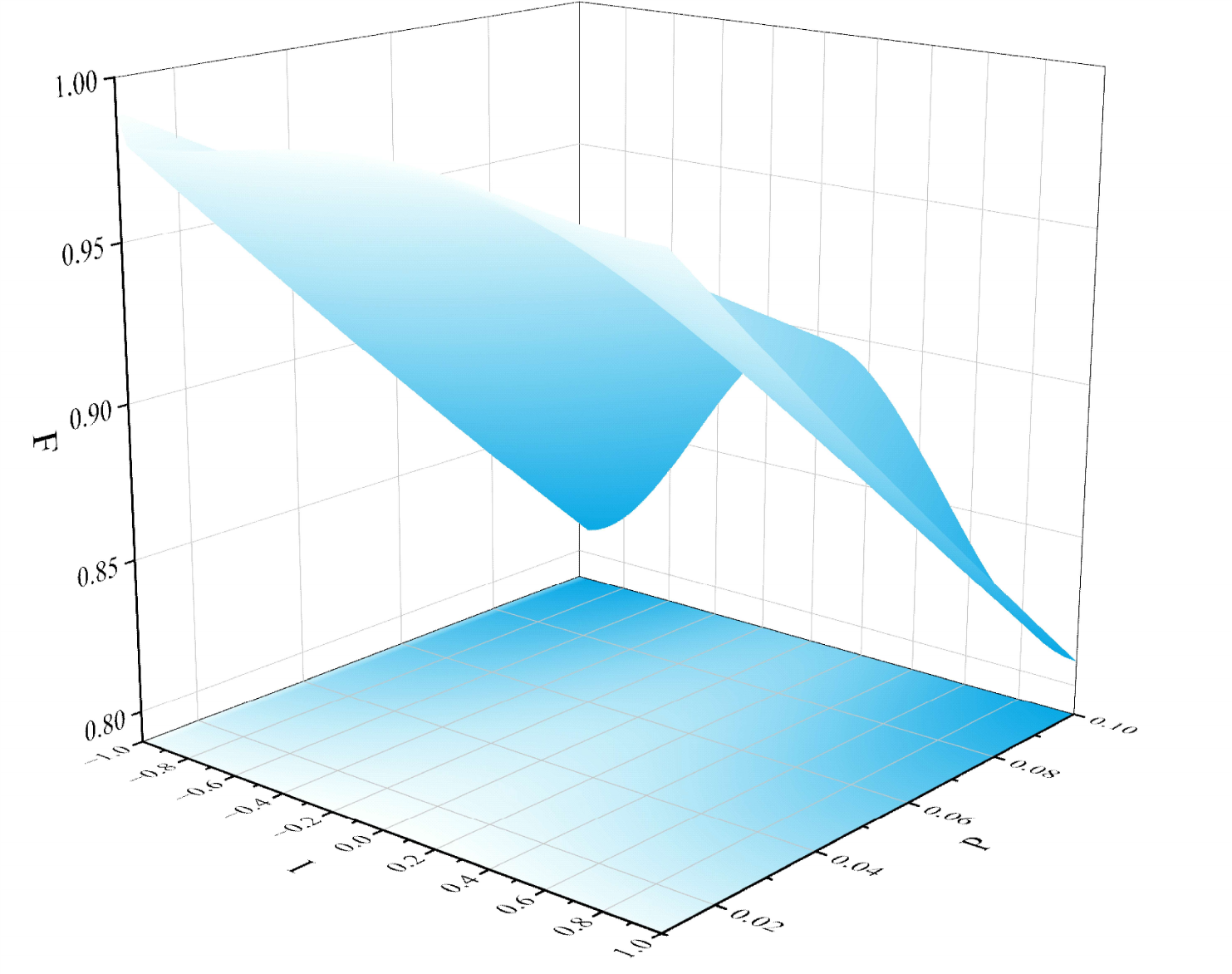}
\label{fig:BF}}
\hfil
\subfloat[Phase Flip]{\includegraphics[width=2.7in]{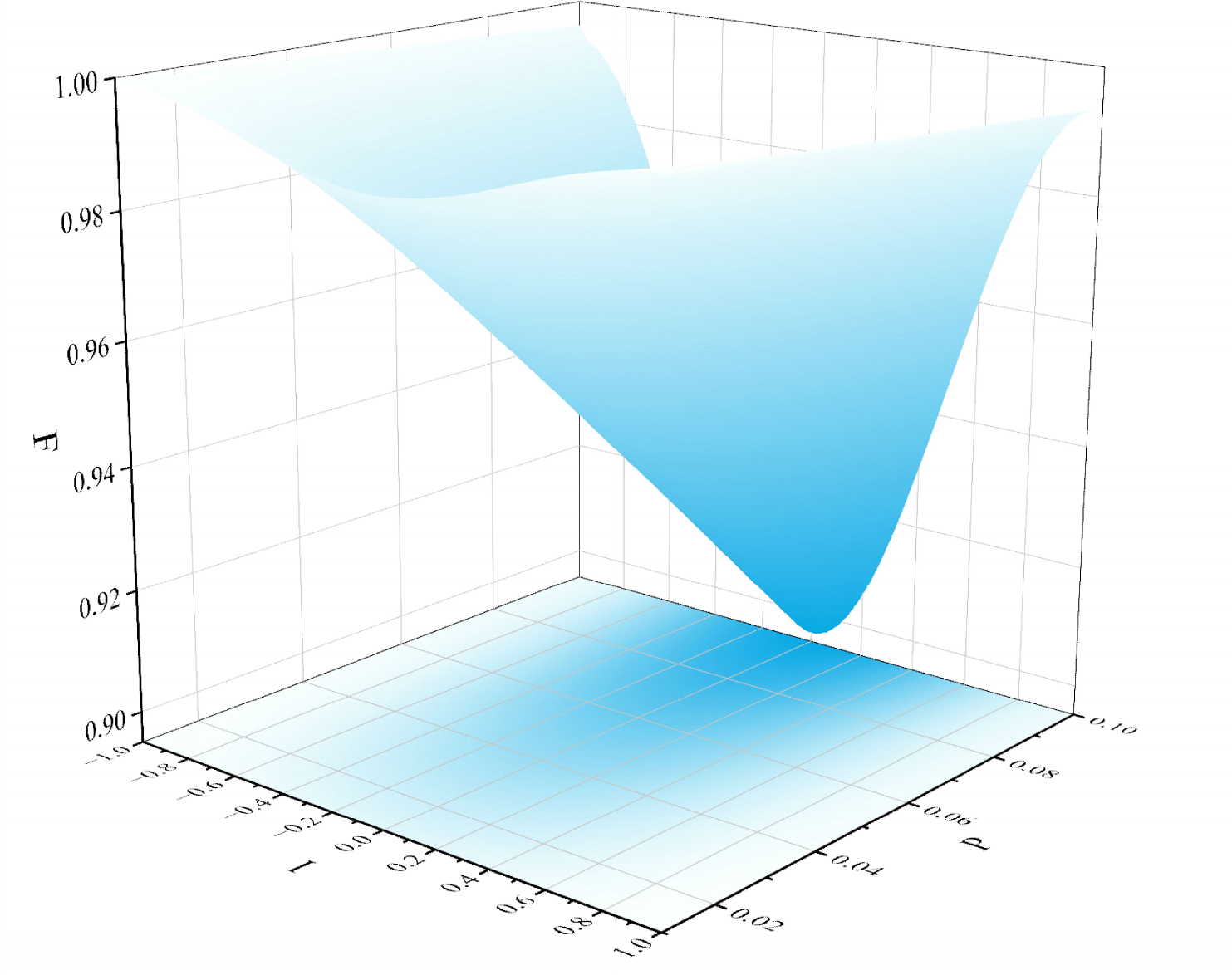}
\label{fig:PF}}
\caption{The fidelity F of the output quantum state, as a function of the input parameter I and the noise probability P, quantifies the similarity between the ideal and noisy states.}
\label{fig:Robustness}
\end{figure*}
When evaluating the robustness of quantum circuits, one crucial factor to consider is their resilience against noise, which often severely impacts circuit performance. Circuits that exhibit poor resistance to noise generally undermine the overall effectiveness of the associated QNN. Thus, ensuring strong noise resilience becomes indispensable when designing robust and reliable quantum circuits.

This section evaluates HQFNN’s resilience by applying four canonical single-qubit noise channels, namely amplitude damping (AD), depolarizing (DP), bit-flip (BF) and phase-flip (PF), to every single-qubit gate with probability \(P\):

\paragraph{Amplitude Damping (AD)}
\begin{align}
	K_0 &= 
	\begin{pmatrix}
		1 & 0\\[6pt]
		0 & \sqrt{1 - P}
	\end{pmatrix},
	&
	K_1 &=
	\begin{pmatrix}
		0 & \sqrt{P}\\[6pt]
		0 & 0
	\end{pmatrix}.
	\label{eq:AD_kraus}
\end{align}

\paragraph{Depolarizing (DP)}
\begin{align}
	K_0 &= \sqrt{1 - \tfrac{3P}{4}}
	\begin{pmatrix}
		1 & 0\\[3pt]
		0 & 1
	\end{pmatrix},\quad
	&
	K_1 &= \sqrt{\tfrac{P}{4}}
	\begin{pmatrix}
		0 & 1\\[3pt]
		1 & 0
	\end{pmatrix},\notag\\
	K_2 &= \sqrt{\tfrac{P}{4}}
	\begin{pmatrix}
		0 & -i\\[3pt]
		i & 0
	\end{pmatrix},\quad
	&
	K_3 &= \sqrt{\tfrac{P}{4}}
	\begin{pmatrix}
		1 & 0\\[3pt]
		0 & -1
	\end{pmatrix}.
	\label{eq:DP_kraus}
\end{align}

\paragraph{Bit Flip (BF)}
\begin{align}
	K_0 &= \sqrt{1 - P}
	\begin{pmatrix}
		1 & 0\\[3pt]
		0 & 1
	\end{pmatrix},
	&
	K_1 &= \sqrt{P}
	\begin{pmatrix}
		0 & 1\\[3pt]
		1 & 0
	\end{pmatrix}.
	\label{eq:BF_kraus}
\end{align}

\paragraph{Phase Flip (PF)}
\begin{align}
	K_0 &= \sqrt{1 - P}
	\begin{pmatrix}
		1 & 0\\[3pt]
		0 & 1
	\end{pmatrix}, 
	&
	K_1 &= \sqrt{P}
	\begin{pmatrix}
		1 & 0\\[3pt]
		0 & -1
	\end{pmatrix}.
	\label{eq:PF_kraus}
\end{align}

To quantify each channel’s effect, we compute the fidelity
\begin{equation}
	F(\rho,\sigma)=Tr\!\Bigl(\sqrt{\sqrt{\rho}\,\sigma\,\sqrt{\rho}}\Bigr)^{2},
\end{equation}
between the ideal output \(\rho\) and the noisy output \(\sigma\), which ranges from 0 (orthogonal) to 1 (identical) \cite{nielsen2010quantum}. Applying the sets \(\{K_i\}\) from \eqref{eq:AD_kraus} to \eqref{eq:PF_kraus} independently to every single‐qubit gate, and sweeping \(P\) from 0.01 to 0.10.

We evaluated the fidelity of HQFNN by sampling the input parameter at equal intervals and report the averages in TABLE \ref{tab:The fidelity on noise} for three noise strengths \(P = 0.01\), 0.05 and 0.10. As \(P\) increases, fidelity falls in each channel but at different rates. Under amplitude damping the fidelity remains above 0.96 even at \(P = 0.10\). With phase flip noise the decline is slightly greater but fidelity still exceeds 0.94 at the highest noise level. Bit flip noise induces a moderate reduction down to about 0.85. Depolarizing noise is comparatively more sensitive and causes the steepest drop as \(P\) increases. Overall, HQFNN maintains high fidelity across all input values and noise channels, with only depolarizing noise exhibiting a slightly more pronounced effect.
As illustrated in Fig.~\ref{fig:Robustness}, each fidelity surface is nearly flat along the input axis, showing that degradation depends almost entirely on noise strength. Under amplitude damping the fidelity remains above 0.96 at \(P = 0.10\). Phase flip noise causes a slightly larger decrease but fidelity still exceeds 0.94 under the heaviest noise. Bit flip noise yields a further drop to approximately 0.85, while depolarizing noise produces the steepest fall as \(P\) increases. These three-dimensional plots underscore HQFNN’s consistently strong robustness across all input values and noise channels.

\subsection{Evaluating Expressibility and Entangling Power}

\begin{table}[htbp]
	\centering
	\caption{Comparison of QNN Expressibility and Entangling Capability.}
	\label{tab:Expressibility and Entangling}
	\begin{tabular}{lccc}
		\toprule
		{\makecell[c]{Fixed\\Parameter}} & {\makecell[c]{Varied\\Parameter}} & Expressibility & {\makecell[c]{Entanglement\\Capacity}} \\
		\midrule
		\multirow{2}{*}{QMF=4}
		& QD=3   & 0.01069 &  0.77506  \\
		& QD=9   & 0.01258 &  0.80444  \\
		\midrule
		\multirow{2}{*}{QD=6}
		& QMF=2  & 0.01616  & 0.83246  \\
		& QMF=6  & 0.01390  & 0.83235  \\
		\midrule
		\multirow{1}{*}{Default}
		& --       & 0.01277  & 0.83551  \\		
		\bottomrule
	\end{tabular}
	\\[3pt]
	{\raggedright\footnotesize ~~QMF: QMF layers. QD: QD qubits. Default: QMF=4, QD=6. \par}
\end{table}

\begin{figure*}[t]
	\centering
	\includegraphics[width=1.8\columnwidth]{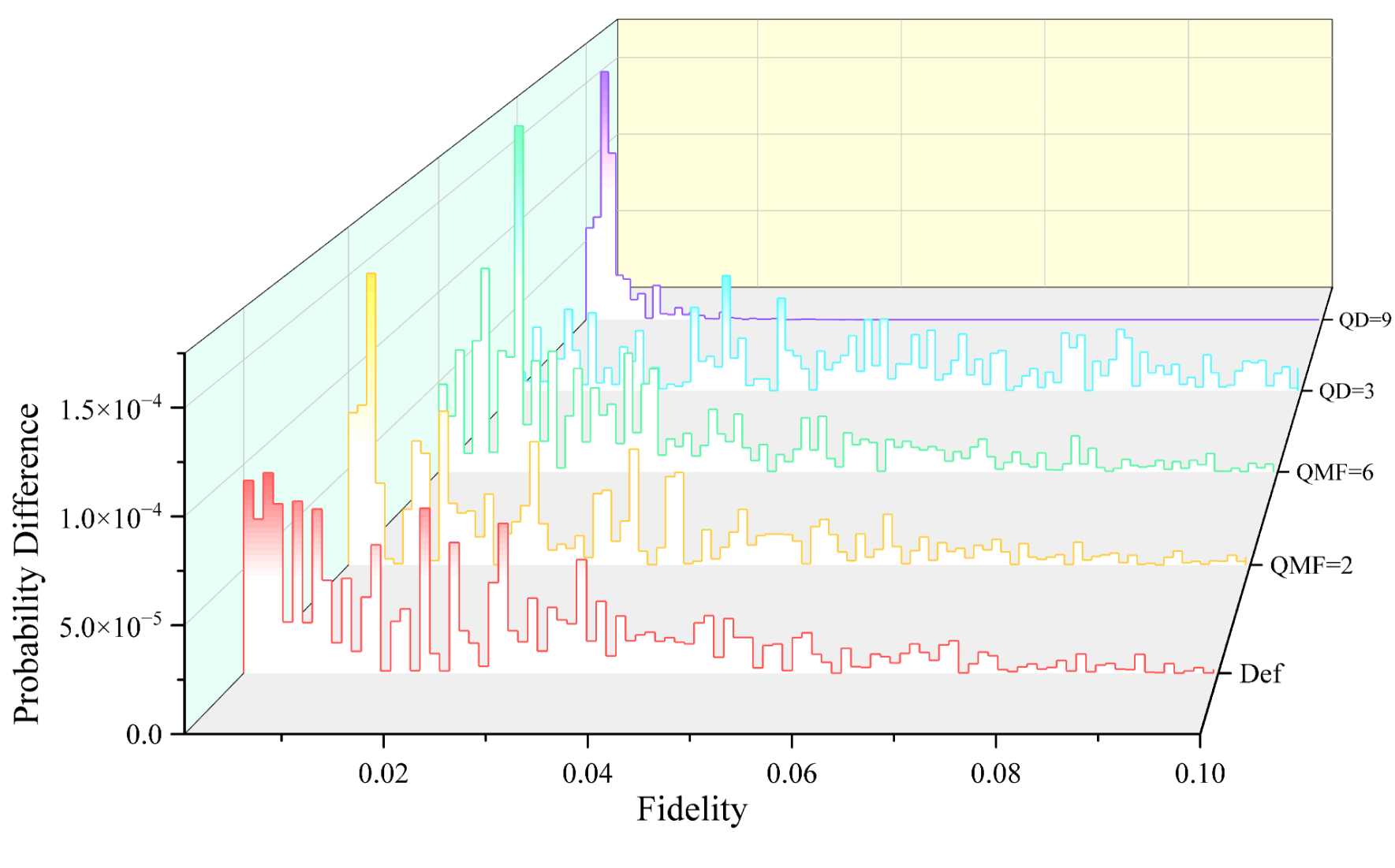}
	\caption{Expressibility comparison of QNN in HQFNN.}
	\label{fig:ExpressibilityAnalysis}
\end{figure*}

Sim et al. \cite{sim2019expressibility} introduced a framework for quantifying a variational circuit’s expressivity and its ability to generate entanglement. The protocol first samples the fidelity distribution between the circuit‐generated states and the ideal Haar‐random states, considered the benchmark for maximal expressiveness, and then computes the Kullback–Leibler divergence between this empirical distribution and the true Haar distribution. The resulting divergence serves as the expressibility score, computed as follows:
\begin{equation}
	\label{eq:quantify expressibility}
	\mathrm{Expr}(\theta)
	= D_{\mathrm{KL}}\bigl(\,\hat{P}_{\mathrm{PQC}}(F;\theta)\,\|\,P_{\mathrm{Haar}}(F)\bigr).
\end{equation}
Here, $F$ represents the fidelity and $\theta$ the adjustable parameters of the PQC.  
$\hat{P}_{\mathrm{PQC}}(F;\theta)$ is the estimated fidelity probability distribution obtained from the PQC block (QMF and QD) in our HQFNN, and  
$P_{\mathrm{Haar}}(F)$ denotes the theoretical fidelity distribution for Haar-random states.  
Thus, a smaller value of $\mathrm{Expr}(\theta)$, i.e.\ lower Kullback-Leibler divergence, means the PQC’s fidelity distribution more closely matches the Haar distribution, indicating greater expressive power of the circuit.

To quantify the circuit’s ability to generate multipartite entanglement, we sample a collection \(S\) of parameter vectors \(\{\theta_i\}\) and compute the Meyer-Wallach measure \(Q\) for each output state \(\lvert\psi_{\theta_i}\rangle\). We then define the average entangling capability as
\begin{equation}
	\mathrm{Ent}=
	\frac{1}{\lvert S\rvert}
	\sum_{\theta_i\in S}
	Q\bigl(\lvert\psi_{\theta_i}\rangle\bigr).
\end{equation}
Here, $Q(\cdot)$ denotes the Meyer–Wallach entanglement measure applied to the $n$-qubit state $\lvert\psi_{\theta_i}\rangle$,  
$S = \{\theta_i\}$ is the set of sampled parameter vectors for the PQC, and  
$\lvert\psi_{\theta_i}\rangle$ is the quantum state prepared by the circuit under parameters $\theta_i$.  
An $\mathrm{Ent}$ value approaching 1 indicates the circuit consistently generates highly entangled states.

Table VI shows a clear trade-off between expressibility and entanglement as we vary one module at a time. 
When QMF depth is held constant, increasing the number of QD qubits first boosts entanglement capacity and then causes it to taper off beyond six qubits, and at the same time expressibility begins to worsen past that point. Although the smallest expressibility appears at three qubits, the accompanying low entanglement makes it an undesirable choice so six qubits strike the best balance and serve as our default. Conversely, when QD size is fixed, entanglement barely changes since it depends almost entirely on the stage containing many CNOT gates, whereas expressibility varies markedly. A very shallow QMF with two layers yields the weakest expressiveness and increasing to six layers actually degrades it again. For this reason, we adopt four QMF layers as the optimal default setting.

Fig.~\ref{fig:ExpressibilityAnalysis}~illustrates the absolute probability difference profiles for various QNN configurations over the fidelity spectrum. For the setting with QD=3, which exhibits the highest expressibility, the fluctuations in the difference curve are notably subdued—a predictable outcome, given its superior ability to approximate the Haar distribution. However, this configuration employs relatively few CNOT gates, yielding lower entanglement and thus engendering potential adverse effects on overall model performance. Meanwhile, the default configuration demonstrates comparably gentle undulations in its difference curve, signifying a closer alignment between QNN‐derived and Haar‐derived probability masses, and such subdued variance is indicative of consistently strong expressibility under the default parameterization.
Notably, when QD increases to 9, the probability difference shifts almost entirely to fidelity values near zero. In this high-dimensional regime, the circuit’s outputs behave like random vectors with virtually no overlap with any single Haar state, so higher‐fidelity bins show negligible probability.


\subsection{Hyperparameter Analysis}

\begin{figure}[t]
	\centering
	\includegraphics[width=1\columnwidth]{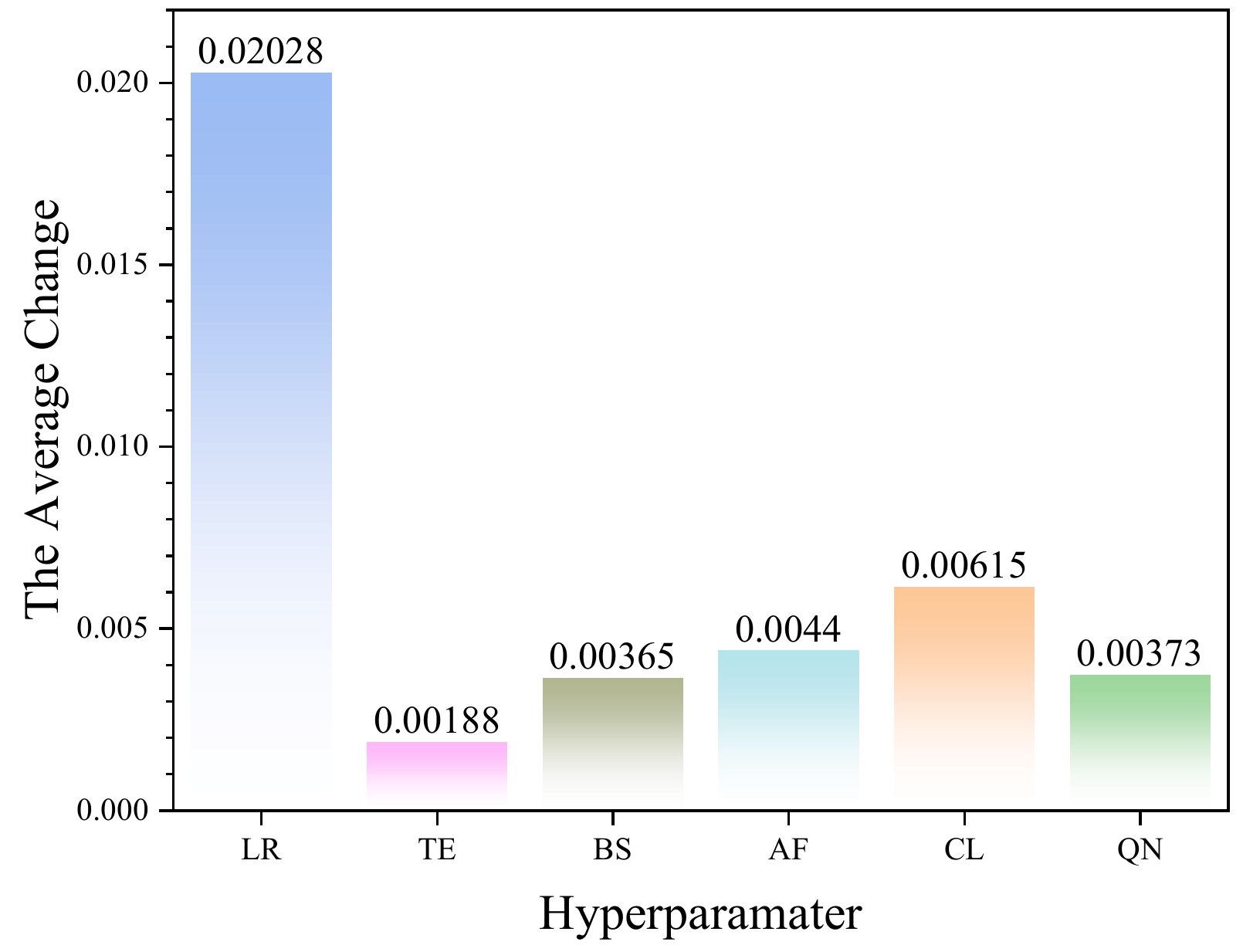}
	\caption{Hyperparameter analysis.}
	\label{fig:Hyperparameter analysis}
\end{figure}

Hyperparameter has non-negligible impact on network performance that is consensus. In this paper, the hyperparameter, like Learning Rate (LR), Training Epochs (TE), Batch Size (BS), Activation Function (AF), the number of Circuit Layers (CL), and Qubits Number (QN) are chosen to implement the sensitivity analysis. Only one parameter is tweaked at a time, in the same time, the other numbers remain at their default values, which are also the configuration values at which the highest performance is obtained. It's necessary to state that according to the characteristics of the proposed model, we designed tests for circuit layers and qubit number, which are used in the QMF and QD modules respectively, and the qubit used in QMF is only one. The complete range or type of parameter are shown in the TABLE \ref{tab:Hyper‑parameter Range}.

Fig.~\ref{fig:Hyperparameter analysis} presents the average change in classification accuracy when each hyperparameter is varied. The learning rate emerges as the single most influential setting, producing an average swing of 2.03\% in model accuracy and outstripping all other candidates by a wide margin. By comparison, increasing the total number of training epochs yields only a 0.19\% shift and enlarging the batch size delivers a modest 0.37\% effect. Adjusting the choice of activation function induces a 0.44\% variation in performance while adding more quantum circuit layers alters accuracy by 0.62\%. Finally changing the qubit count produces the smallest change of 0.37\%. These findings make it clear that effort devoted to tuning the learning rate will yield the most obvious improvement. The activation function has a low influence, indicating that the classical network part accounts for a small proportion of the entire model, which may also indicate that pure quantum fuzzy logic is very worthwhile to develop in image classification tasks.
Once that setting is fixed the model proves remarkably stable across a broad range of epoch budgets, batch sizes, activation schemes, circuit depths and qubit configurations. Consequently we can safely leave most settings near their default values and concentrate on selecting an optimal learning rate schedule.

\begin{table}[t]
	\caption{Hyper-parameter Search Space for HQFNN.}
	\centering
	\begin{tabular}{l l}
		\toprule
		Hyper-parameter  &  Values Explored\\
		\midrule
		Learning rate        & 1$\!\times\!10^{-4}$, 5$\!\times\!10^{-4}$, 5$\!\times\!10^{-3}$, 1$\!\times\!10^{-2}$\\
		Training epochs      & 25, 50, 150, 200             \\
		Batch size           & 100, 300, 700, 900           \\
		Activation function  & Leaky ReLU, Tanh, ELU, GELU  \\
		Circuit layers       & 1, 6, 10                     \\
		Qubits number        & 3, 9, 12                     \\
		\bottomrule
	\end{tabular}
	\label{tab:Hyper‑parameter Range}
	\\[3pt]
	{\raggedright\footnotesize~See Table 1 for default settings.\par}
\end{table}

\subsection{Complexity Analysis}

Complexity is discussed from two viewpoints. Quantum circuit complexity depends on the gate count of the QMF and the QD. Let $d$ be the width of the feature vector after the CNN stem, $m$ the number of QMF rules and $L_{\mathrm q}$ the fixed number of reuploading layers in each rule. A single reuploading layer contains six parameterized rotations, therefore every input sample requires $6L_{\mathrm q} m d$ single-qubit gates.
As the QMF uses one qubit per rule, circuit width stays constant, and the total gate count grows linearly with $m$ and $d$.  The QD operates on a register of $q$ qubits with a constant-depth entangling template, so its gate cost is proportional to $q$ and does not affect the overall scaling.

Computational complexity measures classical arithmetic. The CNN stem maps a mini-batch of $B$ images with $C$ channels and spatial size $H\times W$ at $\mathcal{O}(BCHW)$ cost.  Each of the $B d$ feature components then traverses all $m$ QMF rules, giving $\mathcal{O}(B d m L_{\mathrm q})$ operations.
The rule layer refines the $B\times m\times d$ tensor through a kernel of width $k=3$, adding $\mathcal{O}(B d m k)$ multiplies and $d^{2}k$ weights. The QD projects the tensor to length $q$ and contributes another $\mathcal{O}(B d m)$ operations and $d m q$ parameters, after which a two layer classifier with 64 hidden units introduces $(d+1)64+64C_{\text{cls}}$ weights.  Collecting terms gives end-to-end training complexity $\mathcal{O}(B d m)$ and inference complexity $\mathcal{O}(d m)$, with about $1.6$ M learnable parameters in total, which meets the requirement for a minimal memory footprint and linear computational complexity emphasized in recent quantum-fuzzy studies.



\section{Conclusion}
This work tackles the well-recognized gap between quantum fuzzy reasoning and trainable image classifiers by introducing HQFNN, a highly quantized fuzzy neural network that performs both membership estimation and defuzzification inside parameterized quantum circuits, leaving only a light classical preprocessing step and a linear classifier outside the quantum core. With this arrangement the network learns membership grades directly in Hilbert space, entangles them through a shallow circuit, and blends the resulting quantum representation with a convolutional embedding in a single forward pass. On multiple standard vision benchmarks, the integrated model consistently outperforms classical baselines, fuzzy augmented networks, and quantum only systems, demonstrating the practical benefit of combining quantum superposition with fuzzy logic. Gate count and depth analysis further shows that the quantum component grows gently with image size, keeping computation tractable for higher resolutions. Taken together, the empirical and theoretical results establish HQFNN as both a substantive advance and a flexible platform for continued research in quantum aware fuzzy learning.
Although HQFNN performs respectably on the initial benchmarks, several limitations surface and these in turn shape our next steps. The fixed-depth membership circuit offers only a glimpse of the architecture’s capacity, and a modest expressibility study should show whether extra depth or entanglement truly helps and prevent the gradient flattening noted in pilot runs. Each fuzzy rule is encoded as a single rotation on the Bloch sphere, so a compact geometric look at the learned angles can reveal whether the network captures genuine fuzzy semantics or merely follows pixel statistics, feeding back into lean design adjustments. Present evaluations rely on TorchQuantum’s ideal simulator, and even mild channel noise already trims accuracy, which makes improving robustness through measured gate reordering, gentle parameter tying, or other lightweight error-aware tweaks a natural priority. As these theoretical checks, interpretability probes, and simulator-based noise studies progress, we will scale the pipeline to richer datasets such as CIFAR-100 and ImageNet-mini, aiming to turn HQFNN from a proof of concept into a dependable tool for more demanding vision tasks.

\bibliographystyle{IEEEtran}
\bibliography{Reference}

\begin{thebibliography}{10}
\providecommand{\url}[1]{#1}
\csname url@samestyle\endcsname
\providecommand{\newblock}{\relax}
\providecommand{\bibinfo}[2]{#2}
\providecommand{\BIBentrySTDinterwordspacing}{\spaceskip=0pt\relax}
\providecommand{\BIBentryALTinterwordstretchfactor}{4}
\providecommand{\BIBentryALTinterwordspacing}{\spaceskip=\fontdimen2\font plus
\BIBentryALTinterwordstretchfactor\fontdimen3\font minus
  \fontdimen4\font\relax}
\providecommand{\BIBforeignlanguage}[2]{{%
\expandafter\ifx\csname l@#1\endcsname\relax
\typeout{** WARNING: IEEEtran.bst: No hyphenation pattern has been}%
\typeout{** loaded for the language `#1'. Using the pattern for}%
\typeout{** the default language instead.}%
\else
\language=\csname l@#1\endcsname
\fi
#2}}
\providecommand{\BIBdecl}{\relax}
\BIBdecl

\bibitem{goodfellow2014explaining}
I.~J. Goodfellow, J.~Shlens, and C.~Szegedy, ``Explaining and harnessing
  adversarial examples,'' \emph{arXiv preprint arXiv:1412.6572}, 2014.

\bibitem{mkadry2017towards}
A.~Mkadry, A.~Makelov, L.~Schmidt, D.~Tsipras, and A.~Vladu, ``Towards deep
  learning models resistant to adversarial attacks,'' \emph{stat}, vol. 1050,
  no.~9, 2017.

\bibitem{emde2024towards}
C.~Emde, F.~Pinto, T.~Lukasiewicz, P.~H. Torr, and A.~Bibi, ``Towards
  certification of uncertainty calibration under adversarial attacks,''
  \emph{arXiv preprint arXiv:2405.13922}, 2024.

\bibitem{angelov2010simple}
P.~Angelov and R.~Yager, ``A simple fuzzy rule-based system through vector
  membership and kernel-based granulation,'' in \emph{2010 5th IEEE
  International Conference Intelligent Systems}.\hskip 1em plus 0.5em minus
  0.4em\relax IEEE, 2010, pp. 349--354.

\bibitem{yazdanbakhsh2019deep}
O.~Yazdanbakhsh and S.~Dick, ``A deep neuro-fuzzy network for image
  classification,'' \emph{arXiv preprint arXiv:2001.01686}, 2019.

\bibitem{yeganejou2018classification}
M.~Yeganejou and S.~Dick, ``Classification via deep fuzzy c-means clustering,''
  in \emph{2018 IEEE international conference on fuzzy systems
  (FUZZ-IEEE)}.\hskip 1em plus 0.5em minus 0.4em\relax IEEE, 2018, pp. 1--6.

\bibitem{benedetti2019parameterized}
M.~Benedetti, E.~Lloyd, S.~Sack, and M.~Fiorentini, ``Parameterized quantum
  circuits as machine learning models,'' \emph{Quantum science and technology},
  vol.~4, no.~4, p. 043001, 2019.

\bibitem{havlivcek2019supervised}
V.~Havl{\'\i}{\v{c}}ek, A.~D. C{\'o}rcoles, K.~Temme, A.~W. Harrow, A.~Kandala,
  J.~M. Chow, and J.~M. Gambetta, ``Supervised learning with quantum-enhanced
  feature spaces,'' \emph{Nature}, vol. 567, no. 7747, pp. 209--212, 2019.

\bibitem{schuld2015introduction}
M.~Schuld, I.~Sinayskiy, and F.~Petruccione, ``An introduction to quantum
  machine learning,'' \emph{Contemporary Physics}, vol.~56, no.~2, pp.
  172--185, 2015.

\bibitem{wu2024hierarchical}
S.-Y. Wu, R.-Z. Li, Y.-Q. Song, S.-J. Qin, Q.-Y. Wen, and F.~Gao, ``A
  hierarchical fused quantum fuzzy neural network for image classification,''
  \emph{arXiv preprint arXiv:2403.09318}, 2024.

\bibitem{lowe2004distinctive}
D.~G. Lowe, ``Distinctive image features from scale-invariant keypoints,''
  \emph{International Journal of Computer Vision}, vol.~60, no.~2, pp. 91--110,
  2004.

\bibitem{dalal2005histograms}
N.~Dalal and B.~Triggs, ``Histograms of oriented gradients for human
  detection,'' in \emph{Proceedings of the IEEE Computer Society Conference on
  Computer Vision and Pattern Recognition (CVPR)}, vol.~1, 2005, pp. 886--893.

\bibitem{cortes1995support}
C.~Cortes and V.~Vapnik, ``Support-vector networks,'' \emph{Machine Learning},
  vol.~20, no.~3, pp. 273--297, 1995.

\bibitem{lazebnik2006beyond}
S.~Lazebnik, C.~Schmid, and J.~Ponce, ``Beyond bags of features: Spatial
  pyramid matching for recognizing natural scene categories,''
  \emph{Proceedings of the IEEE Computer Society Conference on Computer Vision
  and Pattern Recognition (CVPR)}, pp. 2169--2178, 2006.

\bibitem{lecun1998gradient}
Y.~LeCun, L.~Bottou, Y.~Bengio, and P.~Haffner, ``Gradient-based learning
  applied to document recognition,'' in \emph{Proceedings of the IEEE},
  vol.~86, no.~11, 1998, pp. 2278--2324.

\bibitem{krizhevsky2012imagenet}
A.~Krizhevsky, I.~Sutskever, and G.~E. Hinton, ``Imagenet classification with
  deep convolutional neural networks,'' in \emph{Advances in Neural Information
  Processing Systems}, vol.~25.\hskip 1em plus 0.5em minus 0.4em\relax Curran
  Associates, Inc., 2012, pp. 1097--1105.

\bibitem{simonyan2015very}
K.~Simonyan and A.~Zisserman, ``Very deep convolutional networks for
  large-scale image recognition,'' in \emph{International Conference on
  Learning Representations (ICLR)}, 2015, arXiv:1409.1556.

\bibitem{szegedy2015going}
C.~Szegedy, W.~Liu, Y.~Jia, P.~Sermanet, S.~E. Reed, D.~Anguelov, D.~Erhan,
  V.~Vanhoucke, and A.~Rabinovich, ``Going deeper with convolutions,'' in
  \emph{Proceedings of the IEEE Conference on Computer Vision and Pattern
  Recognition (CVPR)}, 2015, pp. 1--9.

\bibitem{he2016residual}
K.~He, X.~Zhang, S.~Ren, and J.~Sun, ``Deep residual learning for image
  recognition,'' in \emph{Proceedings of the IEEE Conference on Computer Vision
  and Pattern Recognition (CVPR)}, 2016, pp. 770--778.

\bibitem{tan2019efficientnet}
M.~Tan and Q.~V. Le, ``Efficientnet: Rethinking model scaling for convolutional
  neural networks,'' in \emph{Proceedings of the 36th International Conference
  on Machine Learning (ICML)}.\hskip 1em plus 0.5em minus 0.4em\relax PMLR,
  2019, pp. 6105--6114.

\bibitem{dosovitskiy2021an}
A.~Dosovitskiy, L.~Beyer, A.~Kolesnikov, D.~Weissenborn, X.~Zhai,
  T.~Unterthiner, M.~Dehghani, M.~Minderer, G.~Heigold, S.~Gelly, J.~Uszkoreit,
  and N.~Houlsby, ``An image is worth 16x16 words: Transformers for image
  recognition at scale,'' in \emph{International Conference on Learning
  Representations (ICLR)}, 2021.

\bibitem{liu2022convnext}
Z.~Liu, H.~Mao, C.-Y. Wu, C.~Feichtenhofer, T.~Darrell, and S.~Xie, ``A convnet
  for the 2020s,'' in \emph{Proceedings of the IEEE/CVF Conference on Computer
  Vision and Pattern Recognition (CVPR)}, 2022, pp. 11\,966--11\,976.

\bibitem{zadeh1965fuzzy}
L.~A. Zadeh, ``Fuzzy sets,'' \emph{Information and Control}, vol.~8, no.~3, pp.
  338--353, 1965.

\bibitem{mamdani1975experiment}
E.~H. Mamdani and S.~F. Assilian, ``An experiment in linguistic synthesis with
  a fuzzy logic controller,'' \emph{IEEE Transactions on Computers}, vol. C-24,
  no.~12, pp. 1585--1588, 1975.

\bibitem{klir1995fuzzy}
G.~J. Klir and B.~Yuan, \emph{Fuzzy Sets and Fuzzy Logic: Theory and
  Applications}.\hskip 1em plus 0.5em minus 0.4em\relax Prentice Hall PTR,
  1995.

\bibitem{takagi1985fuzzy}
T.~Takagi and M.~Sugeno, ``Fuzzy identification of systems and its application
  to modeling and control,'' \emph{IEEE Transactions on Systems, Man, and
  Cybernetics}, vol. SMC-15, no.~1, pp. 116--132, 1985.

\bibitem{jang1993anfis}
J.-S.~R. Jang, ``Anfis: Adaptive-network-based fuzzy inference system,''
  \emph{IEEE Transactions on Systems, Man, and Cybernetics}, vol.~23, no.~3,
  pp. 665--685, 1993.

\bibitem{liang2000interval}
Q.~Liang and J.~M. Mendel, ``Interval type-2 fuzzy logic systems: Theory and
  design,'' \emph{IEEE Transactions on Fuzzy Systems}, vol.~8, no.~5, pp.
  535--550, Oct 2000.

\bibitem{mendel2006interval}
J.~M. Mendel, R.~I. John, and F.~Liu, ``Interval type-2 fuzzy logic systems
  made simple,'' \emph{IEEE Transactions on Fuzzy Systems}, vol.~14, no.~6, pp.
  808--824, Dec 2006.

\bibitem{talpur2022deep}
N.~Talpur, S.~J. Abdulkadir, H.~Alhussian, M.~H. Hasan, N.~Aziz, and A.~Bamhdi,
  ``Deep neuro-fuzzy system application trends, challenges, and future
  perspectives: A systematic survey,'' \emph{Artificial Intelligence Review},
  vol.~56, no.~2, pp. 865--913, Apr 2022.

\bibitem{Kak1995}
S.~C. Kak, ``Quantum neural computing,'' \emph{Advances in Imaging and Electron
  Physics}, vol.~94, pp. 259--313, 1995.

\bibitem{Anguita2003}
D.~Anguita, S.~Ridella, F.~Rivieccio, and R.~Zunino, ``Quantum optimization for
  training support vector machines,'' \emph{Neural Networks}, vol.~16, no.
  5--6, pp. 763--770, 2003.

\bibitem{Wiebe2012}
N.~Wiebe, D.~Braun, and S.~Lloyd, ``Quantum algorithm for data fitting,''
  \emph{Physical Review Letters}, vol. 109, no.~5, p. 050505, 2012.

\bibitem{Lloyd2013}
S.~Lloyd, M.~Mohseni, and P.~Rebentrost, ``Quantum algorithms for supervised
  and unsupervised machine learning,'' arXiv preprint arXiv:1307.0411, 2013.

\bibitem{Rebentrost2014}
P.~Rebentrost, M.~Mohseni, and S.~Lloyd, ``Quantum support vector machine for
  big data classification,'' \emph{Physical Review Letters}, vol. 113, no.~13,
  p. 130503, 2014.

\bibitem{Mitarai2018}
K.~Mitarai, M.~Negoro, M.~Kitagawa, and K.~Fujii, ``Quantum circuit learning,''
  \emph{Physical Review A}, vol.~98, no.~3, p. 032309, 2018.

\bibitem{Lloyd2018}
S.~Lloyd and C.~Weedbrook, ``Quantum generative adversarial learning,''
  \emph{Physical Review Letters}, vol. 121, no.~4, p. 040502, 2018.

\bibitem{Cong2019}
I.~Cong, S.~Choi, and M.~D. Lukin, ``Quantum convolutional neural networks,''
  \emph{Nature Physics}, vol.~15, no.~12, pp. 1273--1278, 2019.

\bibitem{Kerenidis2019}
I.~Kerenidis, J.~Landman, A.~Luongo, and A.~Prakash, ``q-means: A quantum
  algorithm for unsupervised machine learning,'' in \emph{Advances in Neural
  Information Processing Systems}, vol.~32, 2019.

\bibitem{sim2019expressibility}
S.~Sim, P.~D. Johnson, and A.~Aspuru-Guzik, ``Expressibility and entangling
  capability of parameterized quantum circuits for hybrid quantum-classical
  algorithms,'' \emph{Advanced Quantum Technologies}, vol.~2, no.~12, p.
  1900070, 2019.

\bibitem{PerezSalinas2020}
A.~Pérez-Salinas, A.~Cervera-Lierta, E.~Gil-Fuster, and J.~I. Latorre, ``Data
  re-uploading for a universal quantum classifier,'' \emph{Quantum}, vol.~4, p.
  226, 2020.

\bibitem{Schuld2020}
M.~Schuld, A.~Bocharov, K.~M. Svore, and N.~Wiebe, ``Circuit-centric quantum
  classifiers,'' \emph{Physical Review A}, vol. 101, no.~3, p. 032308, 2020.

\bibitem{Pesah2021}
A.~Pesah, M.~Cerezo, S.~Wang, T.~Volkoff, A.~T. Sornborger, and P.~J. Coles,
  ``Absence of barren plateaus in quantum convolutional neural networks,''
  \emph{Physical Review X}, vol.~11, no.~4, p. 041011, 2021.

\bibitem{Du2021}
Y.~Du, M.-H. Hsieh, T.~Liu, D.~Tao, and N.~Liu, ``Quantum noise protects
  quantum classifiers against adversaries,'' \emph{Physical Review Research},
  vol.~3, no.~2, p. 023153, 2021.

\bibitem{Qu2022}
Z.~Qu, X.~Liu, and M.~Zheng, ``Temporal-spatial quantum graph convolutional
  neural network based on schrödinger approach for traffic congestion
  prediction,'' \emph{IEEE Transactions on Intelligent Transportation Systems},
  2022.

\bibitem{wangQuantumNASNoiseAdaptiveSearch2022}
H.~Wang, Y.~Ding, J.~Gu, Y.~Lin, D.~Z. Pan, F.~T. Chong, and S.~Han,
  ``{{QuantumNAS}}: {{Noise-Adaptive Search}} for {{Robust Quantum
  Circuits}},'' in \emph{2022 {{IEEE International Symposium}} on
  {{High-Performance Computer Architecture}} ({{HPCA}})}.\hskip 1em plus 0.5em
  minus 0.4em\relax Seoul, Korea, Republic of: IEEE, Apr. 2022, pp. 692--708.

\bibitem{Skolik2023}
A.~Skolik, S.~Mangini, T.~Bäck, C.~Macchiavello, and V.~Dunjko, ``Robustness
  of quantum reinforcement learning under hardware errors,'' \emph{EPJ Quantum
  Technology}, vol.~10, p.~43, 2023.

\bibitem{xiong2025finding}
J.~Xiong, X.-L. Ren, and L.~L{\"u}, ``Finding key nodes in complex networks
  through quantum deep reinforcement learning,'' \emph{Entropy}, vol.~27,
  no.~4, p. 382, 2025.

\bibitem{pykacz1992fuzzy}
J.~Pykacz, ``Fuzzy set ideas in quantum logics,'' \emph{International Journal
  of Theoretical Physics}, vol.~31, no.~9, pp. 1767--1783, 1992.

\bibitem{dallachiara1996fuzzy}
M.~L. Dalla~Chiara and R.~Giuntini, ``Fuzzy quantum logics,'' \emph{Mathware
  and Soft Computing}, vol.~3, no. 1-2, pp. 83--91, 1996.

\bibitem{melnichenko2007quantum}
G.~Melnichenko, ``Quantum decision, quantum logic, and fuzzy sets,''
  \emph{arXiv preprint arXiv:0711.1437}, 2007.

\bibitem{tiwari2024quantum}
P.~Tiwari, L.~Zhang, Z.~Qu, and G.~Muhammad, ``Quantum fuzzy neural network for
  multimodal sentiment and sarcasm detection,'' \emph{Information Fusion}, vol.
  103, p. 102085, 2024.

\bibitem{qu2024quantum}
Z.~Qu, L.~Zhang, and P.~Tiwari, ``Quantum fuzzy federated learning for privacy
  protection in intelligent information processing,'' \emph{IEEE Transactions
  on Fuzzy Systems}, 2024.

\bibitem{wu2024quantum}
S.~Wu, R.~Li, Y.~Song, S.~Qin, Q.~Wen, and F.~Gao, ``Quantum assisted
  hierarchical fuzzy neural network for image classification,'' \emph{IEEE
  Transactions on Fuzzy Systems}, 2024.

\bibitem{jia2025hierarchical}
K.~Jia, F.~Meng, and J.~Liang, ``Hierarchical graph contrastive learning
  framework based on quantum neural networks for sentiment analysis,''
  \emph{Information Sciences}, vol. 690, p. 121543, 2025.

\bibitem{digit-recognizer}
AstroDave and W.~Cukierski, ``Digit recognizer,''
  \url{https://kaggle.com/competitions/digit-recognizer}, 2012, kaggle.

\bibitem{xiao2017fashionmnistnovelimagedataset}
\BIBentryALTinterwordspacing
H.~Xiao, K.~Rasul, and R.~Vollgraf, ``Fashion-mnist: a novel image dataset for
  benchmarking machine learning algorithms,'' 2017. [Online]. Available:
  \url{https://arxiv.org/abs/1708.07747}
\BIBentrySTDinterwordspacing

\bibitem{mukhoti2021deterministic}
J.~Mukhoti, A.~Kirsch, J.~van Amersfoort, P.~H. Torr, and Y.~Gal,
  ``Deterministic neural networks with appropriate inductive biases capture
  epistemic and aleatoric uncertainty,'' \emph{arXiv preprint
  arXiv:2102.11582}, 2021.

\bibitem{lyons2021excavatingaireexcavateddebunking}
\BIBentryALTinterwordspacing
M.~J. Lyons, ``"excavating ai" re-excavated: Debunking a fallacious account of
  the jaffe dataset,'' 2021. [Online]. Available:
  \url{https://arxiv.org/abs/2107.13998}
\BIBentrySTDinterwordspacing

\bibitem{670949}
M.~Lyons, S.~Akamatsu, M.~Kamachi, and J.~Gyoba, ``Coding facial expressions
  with gabor wavelets,'' in \emph{Proceedings Third IEEE International
  Conference on Automatic Face and Gesture Recognition}, 1998, pp. 200--205.

\bibitem{chowdhury2020can}
M.~E. Chowdhury, T.~Rahman, A.~Khandakar, R.~Mazhar, M.~A. Kadir, Z.~B. Mahbub,
  K.~R. Islam, M.~S. Khan, A.~Iqbal, N.~Al~Emadi \emph{et~al.}, ``Can ai help
  in screening viral and covid-19 pneumonia?'' \emph{Ieee Access}, vol.~8, pp.
  132\,665--132\,676, 2020.

\bibitem{nielsen2010quantum}
M.~A. Nielsen and I.~L. Chuang, \emph{Quantum computation and quantum
  information}.\hskip 1em plus 0.5em minus 0.4em\relax Cambridge university
  press, 2010.

\end{thebibliography}
\end{document}